\documentclass[12pt]{article}

\usepackage{graphicx}
\usepackage{amsfonts}
\usepackage{amsmath}
\usepackage{amscd}
\usepackage{latexsym}

\newcommand{\zeetwo}{\mbox{$\mathbb{Z}_{2}$}}
\newcommand{\ket}[1]{\left | #1 \right \rangle}
\newcommand{\bra}[1]{\left \langle #1 \right |}
\newcommand{\amp}[2]{\left \langle #1 \left | #2 \right. \right \rangle}
\newcommand{\ave}[1]{\left \langle #1 \right \rangle}
\newcommand{\proj}[1]{\ket{#1} \! \bra{#1}}

\newcommand{\hilbert}{\mbox{$\mathcal{H}$}}
\newcommand{\absolute}[1]{\left | #1 \right |}

\newcommand{\sys}[1]{^{\mbox{\tiny (#1)}}}

\newcommand{\unity}{\mbox{\bf 1}}
\newcommand{\tr}{\mbox{Tr}\,}
\newcommand{\partr}[1]{\mbox{Tr}_{\mbox{\tiny (#1)}}}
\newcommand{\superop}{\mbox{$\mathcal E$}}
\newcommand{\oper}[1]{#1}

\newcommand{\mket}[1]{\left | #1 \right ) }
\newcommand{\mbra}[1]{\left ( #1 \right | }
\newcommand{\mamp}[2]{\left ( #1 \left | #2 \right. \right ) }

\newcommand{\subsp}[1]{\mathsf{#1}}

\newcommand{\moperation}{\mathcal{E}}
\newcommand{\reduction}[1]{\mbox{\textsf{R}}_{\mbox{\tiny (#1)}}}
\newcommand{\condstate}[2]{\mbox{\textsf{C}} \! \left ( #1 | #2 \right )}

\newcommand{\modalspace}{\mbox{$\mathcal{V}$}}

\newcommand{\dual}[1]{{#1}^{\ast}}

\newcommand{\scalarfield}{\mbox{$\mathcal{F}$}}

\newcommand{\vspan}[1]{\left \langle #1 \right \rangle}

\newcommand{\annihilator}[1]{#1^{\circ}}

\newcommand{\possible}[2]{\mathcal{P} \! \left ( #1 | #2 \right )}
\newcommand{\hvpossible}[2]{\mathcal{P}_{#2} \! \left ( #1 \right )}

\newcommand{\myex}{\mbox{\sf X}}

\title{Almost quantum theory}
\author{Benjamin Schumacher\footnote{Department of Physics, Kenyon College.  Email schumacherb@kenyon.edu} 
    \,\, and Michael D. Westmoreland\footnote{Department of Mathematical Sciences, Denison University.  Email westmoreland@denison.edu}}

\begin{document}

\maketitle

\section{Introductory Remarks}

\subsection{Motivation}

The remarkable features of quantum theory are best appreciated
by comparing the theory to other possible theories---what
Spekkens calls ``foil'' theories \cite{spekkens}.  
The most celebrated example of this approach 
was Bell's analysis \cite{bell}, 
which showed that entangled quantum systems have 
statistical properties unlike any hypothetical local hidden
variable model.  More recently, there have been several
efforts to give quantum theory an operational axiomatic
foundation \cite{hardy,dariano,masanes}.  
In these efforts, a general abstract framework 
is posited to describe system preparations, choices of
measurement, observed results of measurement, and probabilities.
Many possible theories can be expressed in the framework.
The axioms embody fundamental aspects of quantum theory
that uniquely identify it among them.  A striking lesson
of this work is that familiar quantum theory can be
characterized by axioms that seem to have little to do
with the traditional quantum machinery of states and
observables in Hilbert space.  The Hilbert space structure
is ``derived'' from the operational axioms.

These approaches are based on two distinct concepts
of generalization.  First, we generalize within 
quantum theory to give the theory its most general
form.  For example, we generalize state vectors
to density operators as a description of the quantum
state of a system.  Second, we generalize beyond
quantum theory so that we can embed it within a 
wider universe of possible theories.  To be clear,
we refer to these two processes as {\em development}
within a theoretical framework and {\em extension}
beyond that framework.

In this paper, we undertake these processes of 
development and extension, not for actual quantum theory (AQT),
but for a close mathematical cousin of that theory.
Modal quantum theory (MQT) \cite{mqtpaper} is a simplified 
toy model that reproduces many of the structural features of actual 
quantum theory.  The underlying state space of MQT is a vector
space $\modalspace$ over an arbitrary field $\scalarfield$, 
which may be finite.  MQT predicts, not the probabilities
of the results of a measurement, but only which of those
results are possible.  This motivates the use of the term
``modal'', which in formal logic refers to operators
asserting the possibility or necessity of a proposition \cite{modallogic}.
Modal theories themselves can therefore be viewed as
generalizations (extensions) of probabilistic theories.

\subsection{Generalization}  \label{subsec:generalization}

What is ``generalization''?  We begin our
answer to this question with a simple example.  
Suppose we are devising simple
substitution ciphers for English text.  Each letter in
the alphabet $\mathcal{A} = \{A, B, \ldots, Z\}$ is to
be represented by some letter in $\mathcal{A}$.  To 
begin with, we consider only extremely simple ``transposition
ciphers'' in which exactly two letters are exchanged.
For instance, we might exchange $E$ and $R$, leaving
all other letters alone.\footnote{Such a ciphre is not vrey haed
to erad.}  An enciphered message can be decoded by applying
the same transposition a second time.

To generalize this and make better ciphers, we now form 
compound ciphers by applying two or more transposition
ciphers successively.  Enciphered messages are decoded
by applying the same transpositions in reverse order.
Any cipher constructed out of transpositions can 
be described by a {\em permutation} of $\mathcal{A}$, 
an element of $S_\mathcal{A}$.  Furthermore, any
``permutation cipher'' in $S_\mathcal{A}$ can be
constructed in exactly this way, as a compound of
pairwise transpositions.  Thus, the concept of a 
permutation cipher is really a development of the
original idea of a transposition.

Is further development possible?  Consider
the essential requirements for a ``reasonable'' cipher.
A general substitution cipher $c$ is a function 
$c : \mathcal{A} \rightarrow \mathcal{A}$.  Since we
need to be able to recover our plaintext correctly, 
it is appropriate to require as an axiom 
that $c$ be a one-to-one function, so that each
letter in the ciphertext can be decoded in only one way.
Since $\mathcal{A}$ is finite, all one-to-one functions 
on $\mathcal{A}$ are permutations.  This means that the
generalization from transpositions to permutations
already encompasses all reasonable substitution
ciphers (as characterized by our axiom).

To generalize further, we must extend the idea of a cipher
beyond simple letter substitutions.  We might apply
different substitution maps to different letters, or
encipher messages word-by-word.  These more
general ciphers will have some characteristics
in common with substitution ciphers (such as the
unique decipherability of an enciphered message),
but will constitute a larger universe within which
the substitution ciphers form a special class.

In the cipher story we can identify some general features.  
We begin with a basic theory based on a concept $X$.  
The process of development can involve several stages:
\begin{itemize}
\item  {\em Construction.}  In this stage, we devise a
    situation within the existing theory---that
    is, a situation that can be described using $X$---and
    show how this situation can be given a simpler or more
    natural description using $X'$.  ({\em Compositions of
    transposition ciphers can be described as permutation
    ciphers.})
\item  {\em Feasibility.}  Often we are able to show that,
    if a situation is described by $X'$, then it can always
    be given a more cumbersome description in terms of $X$.  
    Informally, every instance of $X'$ is feasible to construct 
    from $X$.  ({\em Every permutation cipher can be described as
    a composition of transposition ciphers.})
\item  {\em Axiomatic characterization.}  We may be led to impose 
    one or more reasonable axioms that any situation 
    ought to satisfy.  Our development is most 
    successful if we can establish that any ``reasonable''
    situation (according to our axioms) can
    be encompassed by our generalized concept $X'$.
    ({\em Any uniquely decodable substitution cipher 
    must be a permutation cipher.})
\end{itemize}
If $X'$ is feasible, then every instance of $X'$ could be 
given a more cumbersome description in terms of $X$.
In this case, the theory including $X'$ is simply a
development of the original one based on $X$.
An axiomatic characterization tells us
that the development is {\em complete}---that no
further reasonable generalization of $X$ is possible 
within the basic theory.

Once we have a complete development from $X$ to $X'$,
further generalization must be an extension of the original
theory.
\begin{itemize}
\item  {\em Extension.}  We can devise a broader framework
    $Y$ within which the theory based on $X$ is a special
    case.  ({\em We can consider ciphers that are not
    based on letter-by-letter substitutions.})
\end{itemize}
Once we have an extended framework $Y$, it is useful to ask
what special properties the original theory may possess.
Thus, we might investigate what distinguishing characteristics
quantum theory has within the wider universe of probabilistic
theories.

\subsection{Scope of the present paper}

The elementary features of modal quantum theory have
been presented elsewhere \cite{mqtpaper,mqtfreewill}.
In the next section, 
we will briefly discuss the axioms for MQT, drawing
the analogies between this theory and AQT.  We will
also discuss some of the properties of entangled 
states of simple systems in MQT.

Following this, 
we turn to a development of MQT analogous to the
standard generalizations of states, measurements and
dynamical evolution in AQT.  Systems whose preparations
are incompletely known, or which are entangled with other
systems, require a more general description of their
states.  Measurement procedures and dynamical
evolution for open systems require additional 
generalizations, which we will also explore.  
As in AQT, we can give
axiomatic characterizations for these new concepts
within the theory, showing that our development is,
in the sense given above, complete.

To generalize further, we must embed MQT within a larger
class of modal theories.  We do this by analogy to
the general probabilistic theories that have been used
to analyze AQT.  As in those theories, our modal theories
are assumed to satisfy a version of the {\em no-signalling 
principle} \cite{gisin}, which states that the choice of 
measurement on one system cannot have a observable effect 
on the measurement results of a distinct system.

Finally, we note that any probabilistic theory can be
viewed through ``modal glasses'', simply interpreting
probabilities $p > 0$ as ``possible'' and $p = 0$
as ``impossible''.  Thus, modal theories are generalizations
of probabilistic theories.  This generalization is
actually an extension, since we will find modal theories
that cannot be ``resolved' to probabilistic ones.  For
situations that arise from systems in MQT,
however, the situation is more complex.  In the
bipartite case we will show that a weak probabilistic
resolution (which may assign $p=0$ for a ``possible''
measurement result) can always be found.

\section{Modal quantum theory}

\subsection{A modal world}  \label{subsec:modalworld}

The world of modal quantum theory is a world without
probabilities.  Probabilities are so familiar that it
is worthwhile to consider more carefully what their
absence entails.

In a probabilistic world, an event $x$ is assigned a
numerical probability $p(x)$ such that $0 \leq p(x) \leq 1$.
The probabilities are normalized, so that 
\begin{equation}
    \sum_{x} p(x) = 1
\end{equation}
where the sum extends over a set of mutually exclusive and
exhaustive events.  Probabilities are related to statistical
frequencies.  Suppose we perform $N$ independent trials of an 
experiment and observe event $x$ to occur $N_{x}$ times.
Then with high probability,\footnote{Note that the connection
between probabilities and statistical frequencies is itself
probabilistic!  This highlights the difficulty in giving a
non-circular operational interpretation of probability.}
\begin{equation}
    p(x) \approx \frac{N_{x}}{N} .
\end{equation}
The possible results of an experiment may be labeled by
numerical values $v$.  The mean of such a random variable
is given by
\begin{equation}
    \ave{v} = \sum_{v} p(v) \, v .
\end{equation}

In a {\em possibilistic} or {\em modal} world, 
we can only distinguish between
possible and impossible events, but we do not assign any
measure of likelihood to them.  That is, we can identify a
possible set
\begin{equation}
    \mathcal{P} = \{ x, x', \ldots \} .
\end{equation}
The only ``normalization'' condition is the requirement that
$\mathcal{P} \neq \emptyset$.  If we perform an experiment many times,
the set $\mathcal{R}$ of results that we see satisfies $R \subseteq P$.
That is, every result we have actually seen is surely 
possible, but we can draw no definite conclusions about
the possibility or impossibility of other results.
Also, without any assignment of ``weights'' to the
numerical results $v$ of an experiment, we cannot compute
a mean value $\ave{v}$.

The naive connection between probabilistic and modal pictures
is that $x \in \mathcal{P}$ if and only if $p(x) \neq 0$.  There are,
however, some subtleties to be recognized.  If we are
assigning probabilities based on observed statistical
frequencies, we cannot distinguish between a very rare
event $x$ (which may not have happened yet in our large but
finite set of trials) and an impossible one.  That is,
we may be able to conclude that $p(x) \approx 0$ but not
that $x$ is impossible.

\subsection{Basic axioms}

The axioms for modal quantum theory are closely related to
those of actual quantum theory, as we can see in
Table~\ref{AQTandMQT}.  The axioms presented are for 
the most elementary versions of each theory; we will
develop them further below.
\begin{table} \label{AQTandMQT}
\begin{tabular}{|p{15em}|p{15em}|}
\multicolumn{1}{c}{\bf Actual quantum theory} & 
\multicolumn{1}{c}{\bf Modal quantum theory} \\[2ex] \hline
{\bf States.}
    A system is described by a Hilbert space $\hilbert$ 
    over the field $\mathbb{C}$ of complex numbers. 
    A state is a normalized $\ket{\psi} \in \hilbert$. &
{\bf States.}
    A system is described by a vector space $\modalspace$
    over a field $\scalarfield$. 
    A state is a non-zero $\mket{\psi} \in \modalspace$. \\ \hline
{\bf Measurements.}
    A measurement is an orthonormal basis $\{ \mket{k} \}$ 
    for $\hilbert$.  Each basis element represents a measurement
    outcome.  For state $\ket{\psi}$, the probability outcome
    $k$ is
    \begin{center} $p(k) = \absolute{\amp{k}{\psi}}^{2}.$ \end{center} &
{\bf Measurements.}
    A measurement is a basis $\{ \mbra{k} \}$
    for $\dual{\modalspace}$.  Each basis element 
    represents a measurement outcome.  For state $\mket{\psi}$,
    outcome $k$ is possible if and only if 
    \begin{center} $\mamp{k}{\psi} \neq 0 .$ \end{center} \\ \hline
{\bf Evolution.}  Over a given time interval, an isolated system
    evolves via a unitary operator $U$:
    \begin{center}$\ket{\psi} \rightarrow U \ket{\psi} .$ \end{center} &
{\bf Evolution.}  Over a given time interval, an isolated system 
    evolves via an invertible operator $T$: 
    \begin{center} $\mket{\psi} \rightarrow T \mket{\psi} .$ \end{center} \\ \hline
{\bf Composite systems.}  The state space for a composite system is
    the tensor product of subsystem spaces:
    \begin{center} $\hilbert\sys{AB} = \hilbert\sys{A} \otimes \hilbert\sys{B} .$ \end{center} & 
{\bf Composite systems.}  The state space for a composite system is
    the tensor product of subsystem spaces:
    \begin{center} $\modalspace\sys{AB} = \modalspace\sys{A} \otimes \modalspace\sys{B} .$ \end{center} \\ \hline
\end{tabular}
\caption{Elementary axioms for AQT and MQT.  \label{table:axioms}}
\end{table}
Even without this development, however, we can identify 
some interesting features of MQT.  Consider,
for instance, the case where $\scalarfield$ is a finite field.
A system with finite-dimensional $\modalspace$ has only a finite
set of possible state vectors.  There are only finitely many
distinct measurements or time evolution maps for the
system, and time evolution must proceed in discrete steps.

The simplest possible system in MQT is a ``modal bit'' or
{\em mobit} \cite{mqtpaper}, for which $\dim \modalspace = 2$.  If we also
choose the smallest field $\scalarfield = \zeetwo$, then
there are just three non-zero vectors in $\modalspace$, which
we can denote $\mket{0}$, $\mket{1}$ and 
$\mket{\sigma} = \mket{0}+\mket{1}$.  The dual space
$\dual{\modalspace}$ also has three vectors, so that
\begin{equation}
    \begin{array}{ccc}
    \mamp{a}{0} = 1 & \mamp{a}{1} = 0 & \mamp{a}{\sigma} = 1 \\
    \mamp{b}{0} = 0 & \mamp{b}{1} = 1 & \mamp{b}{\sigma} = 1 \\
    \mamp{c}{0} = 1 & \mamp{c}{1} = 1 & \mamp{c}{\sigma} = 0 
    \end{array} .
\end{equation}
Any pair of these dual vectors yields a basic measurement.
There are thus three basic mobit measurements corresponding 
to the bases $X = \{ \mbra{c}, \mbra{a} \}$,
$Y = \{ \mbra{b}, \mbra{c} \}$ and
$Z = \{ \mbra{a}, \mbra{b} \}$.  The individual
dual vectors in a measurement basis, which 
correspond to the results of the measurement, are 
called {\em effects}.  It will sometimes be 
convenient to label the measurement results
by $+$ and $-$, so we may write
$\mbra{a} = \mbra{+_{z}} = \mbra{-_{x}}$, etc.
If a mobit is in the state $\mket{\sigma}$ and a
$Z$-measurement is made, both outcomes $+_{z}$ and
$-_{z}$ are possible.

As in AQT, we can compare MQT to a hypothetical
hidden variable theory.  Such a theory supposes that
the system possesses some unknown variable $\lambda$
such that, for a given value of $\lambda$, the
result of any measurement is determined.  As in
AQT, we cannot completely exclude all hidden variable
theories, though we can show that some kinds
of are inconsistent with MQT.  For 
instance, consider a {\em non-contextual} hidden
variable theory \cite{KStheorem}, in which (given $\lambda$) the
question of whether a given effect $\mbra{e}$ 
will occur is independent of which other effects are
present in the measurement basis.  For a given 
value of $\lambda$, the theory would have to
assign ``yes'' or ``no'' values
to each of the dual vectors $\mbra{a}$, $\mbra{b}$
and $\mbra{c}$, such that any pair of them includes
exactly one ``yes''.  This is plainly impossible.
We conclude that the pattern of possibilities for
a mobit system in MQT cannot be reproduced by any
non-contextual hidden variable theory.

This is essentially an MQT version of the famous
Kochen-Specker theorem of AQT \cite{KStheorem}.  
The MQT argument
has a similar structure to the original (both can
be cast as graph-coloring problems) but is 
radically simpler.  Furthermore, the AQT version
of the Kochen-Specker theorem only applies for
$\dim \hilbert \geq 3$, while the MQT version
applies to any system of any dimension \cite{noncontextualityposter}.

\subsection{Entanglement} \label{subsec:entanglement}

Composite systems in MQT may be in either product or
entangled states.  For instance, a pair of $\zeetwo$-mobits
has 15 possible states, of which 9 are product states
and 6 are entangled.  (For more complicated systems,
the entangled states greatly outnumber the product 
states.)

Entangled states are marked by correlated
measurement results.  For example, consider the 
modal ``singlet'' state of two mobits:
\begin{equation}
    \mket{S} = \mket{0,1} - \mket{1,0} .
\end{equation}
(The minus sign here allows us to generalize the state
for any field $\scalarfield$.  For $\scalarfield = \zeetwo$,
$-1 = +1$ and so $\mket{S} = \mket{0,1} + \mket{1,0}$.)
Note that, for any effect $\mbra{e}$,
\begin{equation}
    \mamp{e,e}{S} = \mamp{e}{0} \mamp{e}{1} - \mamp{e}{1} \mamp{e}{0} = 0 .
\end{equation}
Therefore, if we make the same measurement on both mobit
subsystems, it is impossible that we obtain identical
results.

The mobit measurements $X$, $Y$ and $Z$ yield nine possible
joint measurements of a pair of mobits.\footnote{There are
also many measurements involving entangled effects.}
Let $(u,v|U,V)$ denote the situation in which measurements 
of $U$ and $V$ on two systems yield respective results 
$u$ and $v$.  Then we can summarize the measurement results
for $\mket{S}$ as follows:
\begin{itemize}
    \item  If the same measurement is made on each mobit,
        the results must disagree.  Thus $(+,+|X,X)$ is
        impossible, and so on.
    \item  If different measurements are made on the two
        mobits, all but one of the joint results are possible.
        Thus, $(+,-|X,Y)$ is impossible, but $(+,+|X,Y)$,
        $(-,+|X,Y)$ and $(-,-|X,Y)$ are all possible.
\end{itemize}

For AQT, Bell showed that the correlations between entangled 
quantum systems were incompatible with any local hidden 
variable theory \cite{bell}.  He did this by devising a 
statistical inequality that holds for any local hidden
variable theory, but which can be violated by entangled
quantum systems.  Unfortunately, a similar approach based on 
probabilities and expectation values is not available in MQT.

Hardy devised an alternate argument for AQT based only on
possibility and impossibility \cite{hardy}.  He constructed
a non-maximally entangled state $\ket{\Psi}$ for a pair of
qubits together with a set of measurements having 
the following properties:
\begin{itemize}
\item  $(+,+|A,D)$ and $(+,+|B,C)$ are both impossible---that is,
    they have quantum probability $p = 0$.
\item  $(+,+|B,D)$ is possible ($p > 0$).
\item  $(-,-|A,C)$ is impossible ($p = 0$).
\end{itemize}
How might a local hidden variable theory account for this situation?
Since $(+,+|B,D)$ is possible, we restrict our attention to the
set $H$ of hidden variable values that yield this result.
The result of a measurement on one qubit is unaffected by
the choice of measurement on the other (locality).  Furthermore,
no allowed values of the hidden variables can lead to $(+,+|A,D)$ 
or $(+,+|B,C)$.  Thus, for values in $H$, we must obtain
the results $(-,+|A,D)$ and $(+,-|B,C)$.  These jointly imply
that the result $(-,-|A,C)$ would be obtained for values in $H$.
But this contradicts AQT, in which $(-,-|A,C)$ is impossible.

The very same argument applies to the state $\mket{S}$ in MQT,
if we identify $A = X$, $B = Y$, $C = \bar{Z}$ (the negation
of $Z$) and $D = \bar{Y}$.  Thus we can conclude that no local
hidden variable theory can account for the pattern of
possible measurement outcomes generated by the entangled
state $\mket{S}$.

However, this argument has a weakness, because it only applies
to those situations in which the joint outcome $(+,+|B,D) = (+,-|Y,Y)$
actually occurs.  In AQT, we can assign a finite probability $p > 0$
to this result, so we can confidently expect it to arise in a large
enough sample.  But in MQT, the statement that the joint result is
possible does not allow us to draw any such conclusion.  The
MQT version of the Hardy argument therefore applies only to a
situation that may not, in fact, ever occur.

A stronger argument may be constructed along the following 
lines \cite{mqtpaper}.  We imagine that the MQT state $\mket{S}$
corresponds to a set $H_S$ of possible values of a hidden variable.
The variable controls the outcomes of possible measurements in a 
completely local way.
For any particular value $h \in H_S$,
the set of possible results of a measurement on one
mobit depends only on the measurement choice for that
mobit, not on the choice for the other mobit.
Let $\hvpossible{E}{h}$ be the set
of possible results of a measurement of $E$ for the hidden variable
value $h$.  Our locality assumption means that, given $V\sys{1}$ and
$W\sys{2}$ measurements for the two mobits,
\begin{equation}
    \hvpossible{V\sys{A},W\sys{B}}{h} = \hvpossible{V\sys{A}}{h} \times \hvpossible{W\sys{B}}{h},
\end{equation}
the simple Cartesian product of separate sets $\hvpossible{V\sys{1}}{h}$ and
$\hvpossible{W\sys{2}}{h}$.  The MQT set of possible results arising from
$\mket{S}$ should therefore be
\begin{equation}
    \possible{V\sys{1},W\sys{2}}{S} = \bigcup_{h \in H_S}
            \hvpossible{V\sys{1}}{h} \times \hvpossible{W\sys{2}}{h} .
\end{equation}
The individual sets $\hvpossible{V\sys{1}}{h}$, etc., are simultaneously 
defined for all of the measurements that can be made on either mobit.  
Therefore, we may consider the set
\begin{eqnarray}
    \mathcal{J} & = &  \bigcup_{h \in H_S} \,\,
        \hvpossible{X\sys{1}}{h} \times \hvpossible{Y\sys{1}}{h}
        \times \hvpossible{Z\sys{1}}{h} \nonumber \\
        &  &  \qquad {} \times \hvpossible{X\sys{2}}{h}
        \times \hvpossible{Y\sys{2}}{h} \times \hvpossible{Z\sys{2}}{h} .
\end{eqnarray}
There might be up to $2^{6} = 64$ elements in $\mathcal{J}$.  However,
since $\mathcal{J}$ can only contain elements that agree with the
properties of $\mket{S}$, we can eliminate many elements.  For instance,
the fact that corresponding measurements on the two mobits must give
opposite results tells us that $(+,+,+,+,+,+)$ cannot be in $\mathcal{J}$,
though $(+,+,+,-,-,-)$ might be.  However, when we apply all of the
properties of $\mket{S}$ in this way, we find the surprising result
that {\em all} of the elements of $\mathcal{J}$ are eliminated.
{\em No} assignment of definite results to all six possible measurements
can possibly agree with the correspondences obtained from the entangled
MQT state $\mket{S}$.  We therefore conclude that these correspondences
are incompatible with any local hidden variable theory.

This argument can be recast in terms of a {\em pseudo-telepathy
game} \cite{pseudotelepathy}.  Two players, Alice and Bob, are separately 
asked questions drawn from a finite set.  Their goal is to give answers
that satisfy some joint criterion.  The game is a pseudo-telepathy game
if the goal could only be satisfied by classical players if they could
communicate with each other.  However, Alice and Bob may have a winning
strategy if they share quantum entanglement.  In our 
pseudo-telepathy game, Alice and Bob are each asked one of three
questions ($X$, $Y$ or $Z$), and their goal is to provide a joint
answer consistent with the possible measurement outcomes of the
entangled mobit state $\mket{S}$ described above.  If Alice and
Bob answer separately based on shared classical information, they
cannot always win the game.  If they share a mobit pair in $\mket{S}$,
they can.  (However, as we will see below in Section~\ref{Probabilistic resolutions},
this game has no perfect strategy in AQT.)

\section{States and measurements}

\subsection{Generalized states and measurements in AQT}

The axioms for MQT presented in Table~\ref{AQTandMQT} describe
a ``basic'' version of the theory.  In this section and the
next, we will develop the theory to include more general kinds
of states, measurements, and time evolution.  Our development
parallels the standard one in AQT \cite{QPSIbook}, but also has many important
differences.

In AQT, there are situations in which we cannot ascribe a
definite state vector $\ket{\psi}$ to a system, either because
its preparation is not completely known or because we have
a subsystem of a larger composite system in an entangled
state.  In either case, we can construct a description of
the situation from which we can make probabilistic predictions
about the behavior of the system.  We describe such 
{\em mixed states} by means of {\em density operators}.

Suppose, for instance that the system is prepared in one of
several possible pure states, so that $\ket{\psi_{\alpha}}$
occurs with probability $p_\alpha$.  This mixture of states
is described by the density operator
\begin{equation}  \label{rhomixture}
    \rho = \sum_{\alpha} p_{\alpha} \proj{\psi_{\alpha}} .
\end{equation}
If we make a measurement on the system corresponding to
an orthonormal basis $\{\ket{k}\}$, then the overall
probability of the result $k$ is
\begin{equation}
    p(k) = \sum_{\alpha} p_{\alpha} \absolute{\amp{k}{\psi_{\alpha}}}^{2}
        = \bra{k} \rho \ket{k} .
    \label{aqtmixedborn}
\end{equation}
Thus, the density operator $\rho$ is sufficient to predict
the probability of any basic measurement result, given the
probabilistic mixture of states.

Different mixtures of states can yield the same $\rho$,
and thus yield the same statistical predictions.  We
therefore say that the different mixtures correspond to
the same mixed state.  Conversely, different density
operators $\rho$ and $\rho'$ will lead to different
statistical predictions for at least some measurements.

Density operators can also be used to describe a system 
that is part of a composite system.  Given a joint state 
$\ket{\Psi}$ of RQ, we can construct a density operator 
for Q via the partial trace operation:
\begin{equation}  \label{rhopartialtrace}
    \rho = \partr{R} \proj{\Psi} .
\end{equation}
Again, this density operator predicts the probabilities
for any basic measurement on subsystem Q itself, according to
the rule in Equation~\ref{aqtmixedborn}.

Every density operator arising from a mixture or a partial
trace is a positive semidefinite operator of trace 1,
and any such operator could arise in these ways.  The set
of positive semidefinite operators of trace 1 therefore
constitutes our set of generalized states for a system.

We can also develop the concept of measurement in AQT.
As a first step, we can ``coarse-grain'' a basic
measurement, so that each outcome $a$ corresponds to
a projection operator $\Pi_{a}$ (associated with the
subspace of $\hilbert$ spanned by the basis vectors
$\ket{k}$ included in $a$).  We can generalize further
by supposing that we apply our measurement to a composite
system and an {\em ancilla} system, which is regarded as
part of the experimental apparatus.  Then we find that
each outcome $a$ is associated with a positive
semidefinite operator $\oper{E}_{a}$, and that the probability 
of this outcome is 
\begin{equation}
    p(a) = \tr \rho \oper{E}_{a} .  \label{generalbornrule}
\end{equation}
The outcome operators $E_{a}$, sometimes called {\em effect
operators}, sum to the identity:
\begin{equation}
    \sum_{a} \oper{E}_{a} = \unity.  \label{POMsumtounity}
\end{equation}
Our generalized model of measurement is thus a set 
$\{ E_{a} \}$ of positive operators that satisfy 
\ref{POMsumtounity}.  It can be further shown that
any such set can be realized as a coarse-grained
basic measurement on an extended system (i.e., they
are feasible).

Finally, it is possible to 
give an axiomatic characterization of this 
development.  The probability $p(a)$ of measurement 
result $a$ is a functional of $\rho$, more completely written
as $p(a) = p(a|\rho)$.  The state $\rho$ may arise as a
mixture of two other states:
$\rho = p_1 \rho_1 + p_2 \rho_2$.  We infer that the
probability of $a$ for $\rho$ should itself be a probabilistic 
combination:
\begin{equation}  \label{bornrespectsmixtures}
    p(a|\rho) = p_{1} p(a|\rho_{1}) + p_{2} p(a|\rho_{2}) .
\end{equation}
This motivates the axiom that $p(a|\rho)$ is a linear
functional of $\rho$.  Every such linear functional has
the form of Equation~\ref{generalbornrule} for some operator
$E_{a}$.  Since the probability of $a$ must be real and 
non-negative for any state $\rho$, the operator $E_{a}$ 
is positive semidefinite; and since the probabilities 
must always sum to 1, Equation~\ref{POMsumtounity} must 
also hold.

The developments of mixed states and generalized measurements
in AQT thus illustrate the ideas of construction, feasibility
and axiomatic characterization outlined in 
Subsection~\ref{subsec:generalization} above.  We
may regard this as a ``complete'' development within
the theory.  We are now ready to sketch the corresponding
development in MQT.

\subsection{Annihilators and mixed states}

Modal quantum theory is based on a vector space $\modalspace$
of states and its dual $\dual{\modalspace}$ containing 
effects.  It is convenient to summarize here a few 
definitions and elementary results about the subspaces
of $\modalspace$ and $\dual{\modalspace}$ \cite{halmos}.

Subspaces of $\modalspace$ form a lattice under 
the ``meet'' and ``join'' operations $\wedge$ and $\vee$,
where $\subsp{A} \wedge \subsp{B} = \subsp{A} \cap \subsp{B}$ and
$\subsp{A} \vee \subsp{B} = \vspan{\subsp{A} \cup \subsp{B}}$.  
(Here $\vspan{X}$ is the linear span of a set $X$.)  The minimal
subspace in this lattice is $\vspan{0}$, the 
0-dimensional null subspace of $\modalspace$.

Given a set $A$ of vectors in $\modalspace$, the annihilator
$\annihilator{A}$ is the set of dual vectors 
in $\dual{\modalspace}$ that ``annihilate'' vectors in $A$.  
That is,
\begin{equation}
    \annihilator{A} = \left \{ \mbra{e} \in \dual{\modalspace} :
        \mamp{e}{a} = 0 \mbox{ for all } \mket{a} \in A \right \} .
\end{equation}
This can be easily turned around to define 
the annihilator of a subset of the dual space $\dual{\modalspace}$.
In this case, the annihilator would be a subset of
$\modalspace$.

Within modal quantum theory, if $A$ is a set of states, then
$\annihilator{A}$ includes all effects that are impossible
for every state in $A$.  Dually, if $A$ is a set of effects,
then $\annihilator{A}$ includes all states for which every one
of the effects in $A$ is impossible.
In spaces of finite dimension, the annihilator of a set has several 
straightforward properties.
\begin{itemize}
    \item  The annihilator $\annihilator{A}$ is a subspace.
    \item  If $A \subseteq B$, then $\annihilator{B} \subseteq 
        \annihilator{A}$.
    \item  The set $A$ and its span $\vspan{A}$ have
        the same annihilator:  $\annihilator{A} =
        \annihilator{\vspan{A}}$.
    \item  $A$ and $B$ have the same annihilator if and only if
        $\vspan{A} = \vspan{B}$.
    \item $\annihilator{(A \cup B)}
        = \annihilator{A} \wedge \annihilator{B}$.
\end{itemize}
Finally, we note that the annihilator of the annihilator is
a subspace of the original space, and in fact
$A^{\circ \circ} = \vspan{A}$.  If $\subsp{A}$ is a subspace, then
$\subsp{A}^{\circ \circ} = \subsp{A}$.

\subsection{Mixed states in MQT}

In modal quantum theory, a pure state of a system is 
represented by a state vector $\mket{\psi}$ in $\modalspace$.  
How should we represent a mixed state?
We approach this question first by considering mixtures
of pure states.  Since MQT does not involve probabilities, 
a mixture is merely a set of possible state vectors:
$M = \{ \mket{\psi_{1}}, \mket{\psi_{2}}, \ldots \}$.
A particular measurement outcome is possible provided
it is possible for at least one of the states in the 
mixture.  That is, effect $\mbra{e}$ is possible provided
it is {\em not} in the annihilator $\annihilator{M}$.

Two different mixtures $M_{1}$ and $M_{2}$ will 
thus predict exactly the same possible effects if and only if 
$\annihilator{M_{1}} = \annihilator{M_{2}}$, so that 
$\vspan{M_{1}} = \vspan{M_{2}}$.  We say that two such
mixtures yield the same {\em mixed state}, and we identify
that state with the {\em subspace} $\subsp{M} \subseteq \modalspace$ 
spanned by the elements of the mixture.\footnote{This
clarifies a point about pure states, that $\mket{\psi}$ and
$c \mket{\psi}$ are operationally equivalent for any $c \neq 0$.
The two vectors span the same one-dimensional subspace 
of $\modalspace$.}

We can also consider mixtures of two or more
mixed states.  If $\subsp{M}_{1}$ and $\subsp{M}_{2}$ 
are two subspaces of $\modalspace$ associated 
with two states, then $\subsp{M}_1 \vee \subsp{M}_2$
is the subspace associated with a mixture of the two.
Since any non-null subspace of $\modalspace$ can
be written as the span of a set of state vectors, it
is an allowed mixed state.

As in AQT, mixed states in MQT can also arise when 
a composite system is in an entangled pure state.  
Suppose that the composite system RQ is in a 
state $\mket{\Psi\sys{RQ}}$, and
consider the joint effect 
$\mbra{r\sys{R},q\sys{Q}} = \mbra{r\sys{R}} \otimes \mbra{q\sys{Q}}$,
which is possible provided $\mamp{r\sys{R},q\sys{Q}}{\Psi\sys{RQ}} \neq 0$.
We can make sense of this by defining
$\mket{\psi_{r}\sys{Q}} = \mamp{r\sys{R}}{\Psi\sys{RQ}}$.  
That is, if we expand $\mket{\Psi\sys{RQ}}$ in a product basis 
$\{ \mket{a\sys{R},b\sys{Q}} \}$ we can write
\begin{equation} \label{conditionalstate1}
    \mket{\psi_{r}\sys{Q}} 
    = \mbra{r\sys{R}} \left ( \sum_{a,b} \Psi_{ab} \mket{a\sys{R},b\sys{Q}} \right )
    = \sum_{b} \left ( \sum_{a} \Psi_{ab} \mamp{r\sys{R}}{a\sys{R}} \right ) \mket{b\sys{Q}} .
\end{equation}
(The vector $\mket{\psi_{r}\sys{RQ}}$ is independent of the choice of
the $\{ \mket{a\sys{R},b\sys{Q}} \}$ basis chosen for this computation.)
The joint effect $\mbra{r\sys{R},q\sys{Q}}$ is possible provided 
$\mamp{q\sys{Q}}{\psi_{r}\sys{Q}} \neq 0$.  Thus, it makes sense to
interpret $\mket{\psi_{r}\sys{Q}}$ as the conditional state of Q
given the R-effect $\mbra{r\sys{R}}$ for the overall state 
$\mket{\Psi\sys{RQ}}$.\footnote{Of course, if
$\mket{\psi_{r}\sys{Q}} = 0$, then it is not a legitimate state
vector; but in this case, the R-effect $\mket{r}$ is impossible.
The formal inclusion of such phantom conditional states makes
no difference to our analysis.}

To define the unconditional subsystem state for Q, we just
define the Q-subspace of conditional
states for all conceivable R-effects:
\begin{equation}  \label{mqt-partialtrace}
    \subsp{M}\sys{Q} = \{ \mamp{r\sys{R}}{\Psi\sys{RQ}} : \mbra{r\sys{R}} \in \dual{\modalspace\sys{R}} \} .
\end{equation}
An R-measurement is a basis $\{ \mbra{k\sys{R}} \}$ of R-effects.
Since these span $\dual{\modalspace\sys{R}}$, we can see that
\begin{equation}
    \subsp{M}\sys{Q} = \vspan{ \{ \mket{\psi_{k}\sys{Q}} \} },
\end{equation}
where $\mket{\psi_{k}\sys{Q}} = \mamp{k\sys{R}}{\Psi\sys{RQ}}$.

This has the following important consequence.  Whatever measurement
is made on subsystem R, the mixture of conditional Q-states is
exactly $\subsp{M}\sys{Q}$ from Equation~\ref{mqt-partialtrace}.
Thus, the choice of R-measurement by itself makes no observable
difference in the observable properties of Q.

This analysis can be extended to the case where the
composite system itself is in a mixed state.
However, it is more convenient to delay that discussion
until we have also generalized the concept of measurement in MQT.

\subsection{Effects and measurements}

To generalize effects and measurements in modal quantum theory,
it is instructive to begin with an axiomatic characterization.

In the abstract, an effect is simply a map $E$ that assigns 
each subspace $\subsp{M}$ of $\modalspace$ an element of 
$\{ \mbox{possible}, \mbox{impossible} \}$.  But suppose 
$\subsp{M} = \subsp{M}_{1} \vee \subsp{M}_{2}$, the mixture
of states $\subsp{M}_{1}$ and $\subsp{M}_{2}$.  Then 
$E(\subsp{M})$ should be possible if $E$ is possible for
either $\subsp{M}_{1}$ or $\subsp{M}_{2}$.  We therefore
adopt this requirement as an axiom for any reasonable effect 
map $E$.  To make this a consistent rule, we will have to
adopt the sensible convention that $E(\vspan{0})$ is always
impossible.

Our axiom is equivalent to the statement that 
$E(\subsp{M})$ is impossible if and only if both
$E(\subsp{M}_{1})$ and $E(\subsp{M}_{2})$ are
impossible.  Therefore, for a given $E$ we can
consider the subspace $\subsp{Z}_{E} \subseteq \modalspace$ 
\begin{equation}
    \subsp{Z}_{E} = \bigvee \{ \subsp{M} : E(\subsp{M}) \mbox{ is impossible} \} .
\end{equation}
We see that $E(\subsp{M})$ is impossible if and only if 
$\subsp{M}$ is a subspace of $\subsp{Z}_{E}$.  The map $E$ can
therefore be completely characterized by the annihilator
subspace $\annihilator{\subsp{Z}_{E}} \subseteq \dual{\modalspace}$.

A generalized effect in MQT is thus defined to be a
subspace $\subsp{E} \subseteq \dual{\modalspace}$.
For a generalized mixed state $\subsp{M}$, we say that 
$\subsp{E}( \subsp{M} )$ is impossible if $\subsp{M} \subseteq \annihilator{\subsp{E}}$
and possible otherwise.  That is,
\begin{equation}  \label{mqtbornrule1}
\subsp{E}(\subsp{M}) = \left \{
    \begin{array}{lclr}
        \mbox{possible} & \quad & \mamp{e}{m} \neq 0 
            & \quad \mbox{for some } \mbra{e} \in \subsp{E}, \mket{m} \in \subsp{M} \\[2ex]
        \mbox{impossible} & \quad & \mamp{e}{m} = 0 
            & \mbox{for all } \mbra{e} \in \subsp{E}, \mket{m} \in \subsp{M}
    \end{array}
   \right .
\end{equation}
This subspace characterization of generalized effects in MQT is
exactly what we expect from a constructive approach.  Beginning
with an ordinary measurement given by the basis $\{ \mbra{k} \}$ 
for $\dual{\modalspace}$, we can construct a coarse-grained 
effect $\subsp{E}$ from a subset of the $\mbra{k}$ dual vectors.  
This effect is associated with a subspace
of $\dual{\modalspace}$ (the one spanned by the relevant basis vectors).  
Both axiomatic and constructive approaches yield 
the same mathematical representation for generalized effects.

A generalized measurement will be a collection $\{ \subsp{E}_{a} \}$ 
of generalized effects (subspaces of $\dual{\modalspace}$) 
associated with the potential results of the measurement process.  
Some result must always occur, so we impose the requirement that, 
for any state $\subsp{M}$, at least one effect must be possible---that
is, $\subsp{M}$ cannot lie in the annihilator of
all the generalized effects.  Thus,
\begin{equation}
    \bigcap_{a} \annihilator{\subsp{E}_{a}} = \vspan{0},
\end{equation}
and so the generalized effects must satisfy
\begin{equation}
    \bigvee_{a} \subsp{E}_{a} = \dual{\modalspace}.
\end{equation}
This is our ``normalization'' condition for a generalized measurement
in MQT.

In actual quantum theory, a theorem due to Neumark\cite{neumark} 
guarantees that any generalized positive operator measurement on
a system Q can be realized by a basic measurement on a larger
system.  It is not difficult to confirm that an exactly analogous
result holds in MQT.  Thus, our generalized measurements are all
feasible, in the sense discussed in Section~\ref{subsec:generalization}.
The constructive and axiomatic approaches coincide, and so our
development is once again ``complete''.

\subsection{Conditional states} \label{subsec:conditionalstates}

In AQT, any pure entangled state $\ket{\Psi}$ of system RQ
can be written in a special form, the {\em Schmidt 
decomposition} \cite{QPSIbook}, as follows:
\begin{equation}
    \ket{\Psi} = \sum_{k} \sqrt{\lambda_{k}} \,
        \ket{k\sys{R}} \otimes \ket{k\sys{Q}}
\end{equation}
where $\{ \ket{k\sys{R}} \}$ and $\{ \ket{k\sys{Q}} \}$ are
orthonormal bases for the two systems.  It is easy to see
that these are the eigenbases for the subsystem states
$\rho\sys{R}$ and $\rho\sys{Q}$, and that the coefficients
$\lambda_{k}$ are the eigenvalues.  The Schmidt decomposition
is unique except for degeneracy among the $\lambda_{k}$ values
and some choices of relative phases among the two bases.

The MQT analogue of the Schmidt decomposition
can be found as follows.  Suppose $\mket{\Psi\sys{RQ}}$
is a joint state for RQ with subsystem mixed states $\subsp{M}\sys{R}$
and $\subsp{M}\sys{Q}$.  Let $d\sys{R} = \dim \subsp{M}\sys{R}$
and $d\sys{Q} = \dim \subsp{M}\sys{Q}$.
We introduce an R-basis $\{ \mket{k\sys{R}} \}$ for which the first 
$d\sys{R}$ elements form a basis for $\subsp{M}\sys{R}$.  This means
we can write
\begin{equation}
    \mket{\Psi\sys{RQ}} = \sum_{k} \mket{k\sys{R}} \otimes \mket{\psi_{k}\sys{Q}},
\end{equation}
where the sum only requires the first $d\sys{R}$ terms.  The state of
Q is thus $\subsp{M}\sys{Q} = \vspan{\{ \mket{\psi_{k}\sys{Q}} \} }$.
Since $\subsp{M}\sys{Q}$ is spanned by $d\sys{R}$ vectors, we conclude
that $d\sys{R} \geq d\sys{Q}$.  A symmetric argument 
establishes that $d\sys{Q} \geq d\sys{R}$, so the dimensions
are equal.  We therefore identify that $s = d\sys{R} = d\sys{Q}$
as the {\em Schmidt number} of the state $\mket{\Psi\sys{RQ}}$.

The $s$ vectors $\{ \mket{\psi_{k}\sys{Q}} \}$ span a space of 
dimension $s$, so they must be linearly independent.  We can
thus construct a Q-basis $\{ \mket{k}\sys{Q} \}$ in which
$\mket{k\sys{Q}} = \mket{\psi_{k}\sys{Q}}$ for $k \leq s$.  Then
\begin{equation} \label{MQTSchmidtdecomposition}
    \mket{\Psi\sys{RQ}} = \sum_{k} \mket{k\sys{R}} \otimes \mket{k\sys{Q}} ,
\end{equation}
where the sum only includes $s$ terms.  This is a Schmidt
decomposition for $\mket{\Psi\sys{RQ}}$.  It is not unique,
since we had the freedom to choose any basis for the mixed
state (subspace) of one of the systems.

This has a useful consequence.  Given a mixed state $\subsp{M}\sys{Q}$
of Q, an entangled state $\mket{\Psi\sys{RQ}}$ of RQ that leads to
this mixed state is called a {\em purification} of $\subsp{M}\sys{Q}$
in RQ.  Now consider two different purifications $\mket{\Psi\sys{RQ}_{1}}$
and $\mket{\Psi\sys{RQ}_{2}}$ for the same $\subsp{M}\sys{Q}$.  Fixing
a common Q-basis $\{ \mket{k\sys{Q}} \}$, we can write Schmidt
decompositions for both purifications:
\begin{equation}
        \mket{\Psi\sys{RQ}_{1,2}} 
            = \sum_{k} \mket{k\sys{R}} \otimes \mket{k\sys{Q}_{1,2}}.
\end{equation}
The two R-bases are connected by an invertible
operator on $\modalspace\sys{R}$:  $T \mket{k\sys{R}_{1}} 
= \mket{k\sys{R}_{2}}$.  Thus, two purifications of 
$\subsp{M}\sys{Q}$ in RQ are connected via
\begin{equation}
    \mket{\Psi\sys{RQ}_{2}} 
    = \left ( T\sys{R} \otimes \unity\sys{Q} \right ) \mket{\Psi\sys{RQ}_{1}};
    \label{twopurifications}
\end{equation}
that is, by an invertible transformation on R alone.

A theorem of Hughston, Jozsa and Wootters \cite{hjw} (though
earlier discussed by Schr\"{o}dinger \cite{schrodinger} and also by
Jaynes \cite{jaynes}) relates mixtures to entangled states in AQT
and characterizes those mixtures that can give rise to a given 
density operator $\rho$.  An exactly analogous result holds 
in MQT.  First, any mixture for $\subsp{M}\sys{Q}$ 
can be realized as a mixture of conditional states 
arising from a purification of $\subsp{M}\sys{Q}$.  
(The ability to realize different mixtures by a choice of
measurement on the purifying subsystem is the MQT analogue
of the familiar ``steering'' property of actual
quantum theory \cite{steering}.)
Second, the elements of any two mixtures for a given mixed state 
are linear combinations of each other, with coefficients given
by an invertible matrix of scalars.  That is, if 
$\subsp{M} = \vspan{\{ \mket{\psi_{k,1}} \}} = \vspan{\{ \mket{\psi_{k,2}} \}}$,
then 
\begin{equation}
    \mket{\psi_{l,2}} = \sum_{k} T_{lk} \mket{\psi_{k,1}},
\end{equation}
where the $T_{lk}$ form an invertible matrix.

We now return to the question of conditional states and subsystem states
for composite systems in MQT.  How do these ideas work out in the 
context of generalized states and effects?

Suppose the composite system RQ is in the joint state $\subsp{M}\sys{RQ}$, 
and the effect subspace $\subsp{E}\sys{R}$ is part of some 
measurement on R.  The conditional state of Q given this effect, 
which we can denote $\subsp{M}_{E}\sys{Q}$, is defined via a map 
$\condstate{\cdot}{\cdot}$:
\begin{eqnarray} \label{condstatedef}
    \subsp{M}_{E}\sys{Q} & = & \condstate{\subsp{M}\sys{RQ}}{\subsp{E}\sys{R}} \nonumber \\
    & = & \vspan{ \bigg \{  \mamp{e\sys{R}}{m\sys{RQ}} : 
        \mbra{e\sys{R}} \in \subsp{E}\sys{R}, \mket{m\sys{RQ}} \in \subsp{M}\sys{RQ}
             \bigg \} } 
\end{eqnarray}
If $\subsp{M}_{E}\sys{Q} = \vspan{0}$, then the effect $\subsp{E}\sys{R}$ 
is impossible.

The map $\condstate{\cdot}{\cdot}$ respects 
mixtures in both the joint state and the effect.  That is,
\begin{eqnarray}
\condstate{\subsp{M}_{1}\sys{RQ} \vee \subsp{M}_{2}\sys{RQ}}{\subsp{E}\sys{R}} 
    & = & \condstate{\subsp{M}_{1}\sys{RQ}}{\subsp{E}\sys{R}} \vee \condstate{\subsp{M}_{2}\sys{RQ}}{\subsp{E}\sys{R}} \\
\condstate{\subsp{M}\sys{RQ}}{\subsp{E}_{1}\sys{R} \vee \subsp{E}_{2}\sys{R}} 
    & = & \condstate{\subsp{M}\sys{RQ}}{\subsp{E}_{1}\sys{R}} \vee \condstate{\subsp{M}\sys{RQ}}{\subsp{E}_{2}\sys{R}}
\end{eqnarray}
Equation~\ref{condstatedef} generalizes the expression in 
Equation~\ref{conditionalstate1} for conditional states of a
composite system.  It can also be used to define the 
unconditional subsystem state $\subsp{M}\sys{Q}$, which we denote
like so:
\begin{equation}\label{modalrelativestatereduction}
\subsp{M}\sys{Q} = \reduction{R} \left ( \subsp{M}\sys{RQ} \right ) 
    = \condstate{\subsp{M}\sys{RQ}}{\dual{\modalspace\sys{R}}} .
\end{equation}
Since we take the linear span in Equation~\ref{condstatedef},
we only need to consider spanning sets for
$\subsp{M}\sys{RQ}$ and $\subsp{E}\sys{R}$.  That is, if
$\subsp{M}\sys{RQ} = \vspan{\{ \mket{\mu\sys{RQ}} \}}$ and
$\subsp{E}\sys{R} = \vspan{\{ \mbra{\eta\sys{R}} \}}$, then
$\subsp{M}_{E}\sys{Q} = \vspan{ \{ \mamp{\eta\sys{R}}{\mu\sys{RQ}} \} }$.
This is useful in calculations.

\section{Open system evolution}

\subsection{Type M maps}

According to the evolution postulates given in Table~\ref{table:axioms},
the time evolution of state vectors in either actual or modal quantum
theory can be described by a linear operator---unitary in the case of AQT
($\ket{\psi} \rightarrow U \ket{\psi}$), invertible in the case
of MQT ($\mket{\phi} \rightarrow T \mket{\phi}$).  In either case 
it is straightforward to generalize this to mixed states.  The density 
operator in AQT evolves via $\rho \rightarrow U \rho U^{\dagger}$, and
in MQT a subspace evolves according to
\begin{equation}
    \subsp{M} \rightarrow T \subsp{M} 
        = \left \{ T \mket{\phi} : \mket{\phi} \in \subsp{M} \right \} .
\end{equation}
These postulates apply when the system in question is isolated.  
When a system is subject to noise or interaction with its
environment, a more general description of time evolution 
is needed.  In this section we trace this development.

Generalized operations can be of two types.  {\em Conditional}
operations do not take place with certainty but only happen
when some objective condition (e.g., a measurement result)
is observed.  {\em Unconditional} operations are those that
take place with certainty.

In actual quantum theory, a general operation is a map on 
density operators:  $    \rho \rightarrow \superop (\rho)$.
For an input density operator $\rho$, the output
$\superop(\rho)$ of an unconditional operation must
also be a density operator---that is, a positive 
semidefinite operator of trace 1.  For conditional
operations, the output is subnormalized so that
$p = \tr \superop(\rho)$ is the probability that 
the operation occurs.
The map $\superop$ must respect mixtures; that is,
\begin{equation}
    \superop \left ( p_{1} \rho_{1} + p_{2} \rho_{2} \right )
        = p_{1} \superop(\rho_{1}) + p_{2} \superop ( \rho_{2} ) .
\end{equation}
Thus $\superop$ is a linear map on density operators.
This is a powerful condition, since it allows us to extend
$\superop$ to a linear map on the space of all operators---that is, 
to a {\em superoperator}.

In MQT, a general operation $\moperation$ on a system is 
a map on the subspaces of $\modalspace$:
$\subsp{M} \rightarrow \subsp{M}' = \moperation \left ( \subsp{M} \right )$.
For an unconditional operation, the output of the map must
always be a legitimate state, a non-null subspace.
This means that $\moperation (\subsp{M}) \neq \vspan{0}$ for 
$\subsp{M} \neq \vspan{0}$.  This requirement is relaxed for 
conditional operations.  In that case, 
$\moperation (\subsp{M}) = \vspan{0}$
merely signifies that the condition of the operation cannot
arise for the input state $\subsp{M}$.

General operations in MQT must also respect mixtures,
meaning that
\begin{equation}  \label{typeMcondition}
    \moperation \left ( \subsp{M}_{1} \vee \subsp{M}_{2} \right )
        = \moperation ( \subsp{M}_{1} ) \vee \moperation ( \subsp{M}_{2} ).
\end{equation}
(To maintain consistency, we adopt the convention that 
$\moperation ( \vspan{0} ) = \vspan{0}$.)
The map $\moperation$ is not simply a linear superoperator, so
we cannot easily extend it to inputs other than subspaces.
However, Equation~\ref{typeMcondition} is still an important 
requirement.  We will call subspace maps that respect mixtures
in this way {\em Type M} maps.

Throughout the rest of this section, we will only consider 
unconditional operations in both AQT and MQT.  The 
generalization to conditional operations is not difficult
and is left as an exercise.

\subsection{Constructive approach}

Consider a situation in which the system of interest $S$ interacts 
with an external ``environment'' system $E$, where $E$ is initially 
in some fixed state. In AQT, the dynamics  of just such an {\em open} 
system $S$ is described by a map $\superop$ on density operators:
\begin{equation}\label{unitarysuperop}
    \rho \rightarrow \superop (\rho) =
        \partr{E} \oper{U} \left ( \rho \otimes
        \proj{0} \right ) \oper{U}^{\dagger}
\end{equation}
where $\ket{0}$ is the initial standard state of $E$ and $\oper{U}$ 
is a unitary operator on the composite system SE.

By analogy, in MQT the evolution of an open system that interacts
with an environment (initial state $\subsp{M}_{0}\sys{E})$ can
be described by the map ${\superop}\sys{S}$ such that
\begin{equation}
     {\superop}\sys{S}({\subsp{M}}\sys{S}) = \reduction{E}(T\sys{SE}({\subsp{M}}\sys{S}\otimes {\subsp{M_0}}\sys{E}))
\end{equation}                  
where $\reduction{E}$ is the subsystem state reduction
defined by equation (\ref{modalrelativestatereduction}). 
Without loss of generality, we may suppose that $\subsp{M}_{0}\sys{E}$
is one-dimensional, since any mixed environment state can have a
purification in a larger environment.
We refer to these maps defined by invertible linear evolution on a  
larger system as {\em Type I} maps.

In AQT there is a way 
of representing the map $\superop$ without the explicit involvement
of the environment E in Equation~\ref{unitarysuperop}.
Consider a particular basis $\{ \ket{e_{k}} \}$
for the Hilbert space of the environment E.  
For each $k$ define the operator $\oper{A}_{k}$ by
\begin{equation}
    \oper{A}_{k} \ket{\phi} = \bra{e_{k}} \oper{U} \ket{\phi,0} 
    \label{krausoperatordef}
\end{equation}
for any $\ket{\phi}$ in $\hilbert\sys{S}$.  Even though we have
used the environment E and the interaction $\oper{U}$ in this
definition, the $\oper{A}_{k}$ operators  act on 
$\hilbert\sys{S}$ alone.  We may use the $\{ \ket{e_{k}} \}$ 
basis to do the partial trace in Equation~\ref{unitarysuperop}.
Given a pure state input $\ket{\phi}$,
\begin{eqnarray}
    \superop ( \proj{\phi} ) 
    & = & \sum_{k} \bra{e_{k}} \oper{U} \left ( \proj{\phi}
        \otimes \proj{0} \right ) \oper{U}^{\dagger} \ket{e_{k}} \nonumber \\
    & = & \sum_{k} \oper{A}_{k} \proj{\phi} \oper{A}_{k}^{\dagger} .
\end{eqnarray}
And in general,
\begin{equation}
    \superop ( {\rho} ) = \sum_{k} \oper{A}_{k} {\rho} 
        \oper{A}_{k}^{\dagger} .
    \label{operatorsumrep}
\end{equation}
This is called an {\em operator-sum representation} or {\em Kraus
representation} of the map $\superop$, and the operators $\oper{A}_{k}$
are called {\em Kraus operators} \cite{QPSIbook}.

For an unconditional operation, 
the Kraus operators satisfy a normalization condition.  
If ${\rho}' = \superop(\proj{\phi})$ for a normalized
pure state $\ket{\phi}$, then
\begin{eqnarray}
    \tr {\rho}' 
    & = & \sum_{k} \bra{\phi} \oper{A}_{k}^{\dagger} \oper{A}_{k} 
        \ket{\phi} \nonumber \\
    & = & \bra{\phi} \left ( \sum_{k} \oper{A}_{k}^{\dagger}
        \oper{A}_{k} \right ) \ket{\phi} .
\end{eqnarray}
Since $\tr {\rho}' = 1$ for every normalized input state
$\ket{\phi}$, 
\begin{equation}
    \sum_{k} \oper{A}_{k}^{\dagger} \oper{A}_{k} = \unity .
    \label{operatorsumnormalization}
\end{equation}
We can describe the map $\superop$ entirely in 
terms of Kraus operators.  It is often much
more convenient to describe $\superop$ in this way without
considering the actual environment E, which might be 
very large and complex.  An operator-sum representation 
is a compact description of how E affects the evolution 
of the state of S.

We can make the analogous construction in MQT.  The system
S interacts with an environment E, initially in the state
$\subsp{M}_{0}\sys{E} = \vspan{\mket{0\sys{E}}}$, via the
invertible operator $T\sys{SE}$.
Let $\{ \mbra{e_{k}\sys{E}} \}$
be a basis for $\dual{\modalspace\sys{E}}$ and define the
operator $A_{k}$ on $\modalspace\sys{S}$ by
\begin{equation}
    A_{k} \mket{\phi\sys{S}} = \mbra{e_{k}\sys{E}} T\sys{SE}
        \mket{\phi\sys{S}, 0\sys{E}} .
\end{equation}
Given an initial S-state $\subsp{M}\sys{S}$ spanned by state
vectors $\mket{m\sys{S}}$, we have
\begin{eqnarray}
    \superop\sys{S} \left ( \subsp{M}\sys{S} \right )
    & = & \reduction{E}(T\sys{SE}({\subsp{M}}\sys{S}\otimes {\subsp{M_0}}\sys{E})) \nonumber \\
    & = & \vspan{ \mbra{e_{k}\sys{E}} T\sys{SE} \mket{m\sys{S}, 0\sys{E}} } \nonumber \\
    & = & \vspan{ A_{k} \mket{m\sys{S}} } \nonumber \\
    \superop\sys{S} \left ( \subsp{M}\sys{S} \right )
    & = & \bigvee_k A_k\subsp{M}\sys{E} . \label{MQTKrausrep}
\end{eqnarray}
The output of $\superop\sys{S}$ acting on $\subsp{M}\sys{S}$
is a mixture of images of $\subsp{M}\sys{S}$ under the linear
operators $A_{k}$.  Equation~\ref{MQTKrausrep} is the MQT analogue 
of the Kraus representation in Equation~\ref{operatorsumrep}.  
If a map $\superop\sys{S}$ has a representation of this type,
we say that it is a {\em Type L} map.  (Note that we have shown 
that all Type I maps are also Type L.)

The individual operators $A_{k}$ are not necessarily invertible.
However, if $\superop$ represents an unconditional
operation on the MQT system S, then any non-null subspace $\subsp{M}$
must evolve to a non-null subspace $\superop (\subsp{M})$.  
Thus, the $A_{k}$ operators must satisfy
\begin{equation} \label{MQToperatorsumnormalization}
\bigcap_k \ker A_k =  \langle 0 \rangle ,
\end{equation}
the MQT analogue of the normalization condition in
Equation~\ref{operatorsumnormalization}.

\subsection{Axiomatic characterization}

In AQT, every ``physically reasonable'' dynamical evolution map for
an open system has both a unitary and a Kraus representation.
Similarly, every ``physically reasonable'' evolution map in MQT
is both Type I and Type L.  To make sense of this claim, we must
explain what is meant by a ``physically reasonable'' map.

Let us begin by reviewing the argument in AQT.
A physically reasonable map $\superop$ must be linear in 
the input density operator, making it a superoperator (an
element of $\cal{B}(\cal{B}(\hilbert))$).  Furthermore,
the output of $\superop$ must be a valid density operator 
for any valid input state.  This immediately implies two 
properties:
\begin{itemize}
    \item  $\superop$ must be a {\em positive} map, in the sense 
        that it maps positive operators to positive operators.
    \item  $\superop$ must be a {\em trace-preserving} map, so
        that $\tr \superop(\oper{A}) = \tr \oper{A}$ for all
        operators $\oper{A}$.
\end{itemize}
(Both properties are easy to prove in general, since any operator 
can be written as a linear combination of density operators.)

These two conditions are not sufficient to characterize 
``physically reasonable'' linear maps, because there are
positive, trace-preserving maps that cannot correspond to the
time evolution of an quantum system.  The easiest example
arises for a simple qubit system.  The Pauli operators
$\oper{X}$, $\oper{Y}$ and $\oper{Z}$, together with the
identity $\unity$, form an operator basis.  The following
map $\mathcal{T}$ is positive:
\begin{equation}
    \begin{array}{ccc}
    \mathcal{T}(\unity) = \unity & \quad & \mathcal{T}(\oper{X}) = \oper{X} \\
    \mathcal{T}(\oper{Y}) = \oper{Y} & & \mathcal{T}(\oper{Z}) = - \oper{Z}
    \end{array} .
\end{equation}
However, $\mathcal{T}$ does not describe the possible evolution of
the state of an open qubit system.  The reason is that the qubit
is not necessarily alone in the universe.  We may consider a second
independent qubit whose state
evolves according to the identity map $\mathcal{I}$.
The composite system evolves according to the map $\mathcal{T} \otimes \mathcal{I}$.
However, this extended map is {\em not} positive:  entangled input
states may map to operators having some negative eigenvalues.
(See \cite{QPSIbook} for details.)

We need stronger property to characterize ``physically reasonable''
evolution maps for an open quantum system.  The structure of the
example just described provides a clue to what this stronger
property looks like.

The map $\superop$ for a system is said to be {\em completely 
positive} if $\superop \otimes \mathcal{I}$ is positive 
whenever we append an independent quantum system.
This means that, for any initial pure state $\ket{\Psi}$ of the
composite system, the operator $\superop \otimes \mathcal{I}
( \proj{\Psi} )$ is positive.  Since any system may be part of
a composite system, we require that every ``physically reasonable''
map describing open system evolution must be linear, trace-preserving
and completely positive.
The importance of this requirement is shown by the following theorem.
\begin{quotation}
    \noindent
    {\bf Representation theorem for generalized dynamics in AQT.}  
    Let Q be a quantum system and $\superop$ be a map on 
    Q-operators.  The following conditions are equivalent.
    \begin{description}
        \item[(a)]  $\superop$ is a linear, trace-preserving,
            completely positive map.
        \item[(b)]  $\superop$ has a ``unitary representation''.
            That is, we can introduce an environment system E,
            an initial environment state $\ket{0}$ and a joint
            unitary evolution $\oper{U}$ on QE so that
            \begin{equation}
            \superop ( \oper{G} ) = \partr{E} \oper{U} \left ( \oper{G}
                \otimes \proj{0} \right ) \oper{U}^{\dagger} .
            \end{equation}
        \item[(c)]  $\superop$ has a Kraus representation.  That is,
            we can find operators $\oper{A}_{k}$ such that
            \begin{equation}
            \superop ( \oper{G} ) = \sum_{k} \oper{A}_{k} \oper{G} 
            \oper{A}_{k}^{\dagger} .
            \end{equation}
            The Kraus operators satisfy the normalization condition
            of Equation~\ref{operatorsumnormalization}.
     \end{description}
\end{quotation}
Once again, the constructive approach (unitary dynamics on a
larger system) and the axiomatic characterization (linear, 
trace-preserving, completely positive maps) lead us to the
same generalized unconditional operations in AQT.  
The representation theorem is a powerful and fundamental result in AQT.
Details of its proof are given in Appendix D of \cite{QPSIbook}.

So much for AQT.  Can we find an analogous axiomatic characterization
for ``reasonable'' state evolution in MQT?  This is a tricky
question, and not just because we lack access to
an actual MQT world.
A proposed evolution map $\superop$ will map subspaces to 
subspaces, rather than operators to operators.  Furthermore,
the underlying field $\scalarfield$ 
may not include the notion of ``positive elements''.  
In a field of non-zero characteristic,
any element added to itself sufficiently many times yields 0.
Thus, in MQT there may be no analogue to the notion of a 
``positive map''.

Nevertheless, there is a close analogue to the property of complete
positivity that does make sense in MQT.  As in AQT, this condition
governs how a map $\superop$ extends to one that applies to
a larger composite system.  Briefly, we require that an extension
exists that commutes with the conditioning operation described
in Subsection~\ref{subsec:conditionalstates}.  Here is a more
precise definition: We say that the subspace map $\superop\sys{S}$ 
for modal quantum system S is {\em Type E} if for any other 
system R there exists a joint subspace map $\superop\sys{RS}$ such that
\begin{equation}
  \superop\sys{S}( \condstate{\subsp{M}\sys{RS}}{\subsp{E}\sys{R}}) =
         \condstate{\superop\sys{RS}(\subsp{M}\sys{RS})}{\subsp{E}\sys{R}} 
\end{equation}
for any RQ-state $\subsp{M}\sys{RQ}$ and R-effect $\subsp{E}\sys{R}$.  
In other words, for a Type E map $\superop\sys{S}$ and a system R, 
we can find a map $\superop\sys{RS}$ so that the following 
diagram always commutes:
\begin{equation}  \label{smallcommute}
    \begin{CD}
       \subsp{M}\sys{RS} @>\mbox{$\condstate{\cdot}{\subsp{E}\sys{R}}$}>> \subsp{M}\sys{S}
            \\[2ex]
        @V \superop\sys{RS} VV  @VV \superop\sys{S} V
            \\[2ex]
         {\superop\sys{RS}(\subsp{M}\sys{RS})} @> \mbox{$\condstate{\cdot}{\subsp{E}\sys{R}}$} >> 
             \superop\sys{S}(\subsp{M}\sys{S})
    \end{CD}
\end{equation}
We will require that any ``reasonable'' state evolution in MQT
must be Type E.  What is the motivation for such a condition?
Suppose the state of system S evolves according to $\superop\sys{S}$.
It is always reasonable to suppose that another system R exists
in the MQT universe.  We imagine that R and S can be ``independent''
of one another---they might, for instance, be very
far apart in space.  The joint system RS is initially in the state 
$\subsp{M}\sys{RS}$.  The evolution of RS is described by some 
joint map $\superop\sys{RS}$ that reflects the 
independence of the subsystems.  
We now imagine two experimental procedures.
\begin{itemize}
    \item  A measurement is made on system R, with the objective
        result corresponding to effect $\subsp{E}\sys{R}$.
        Under this condition, S is in the state 
        $\condstate{\subsp{M}\sys{RS}}{\subsp{E}\sys{R}}$.
        Now the dynamical evolution acts, so that the final
        state of S is
        $\superop\sys{S} \left ( \condstate{\subsp{M}\sys{RS}}{{\subsp{E}\sys{R}}} \right )$.
    \item  The dynamical evolution acts, leading to the joint
        final state $\superop\sys{RS}(\subsp{M}\sys{RS})$.  Now
        the measurement is made on system R, with the objective
        result corresponding to the effect $\subsp{E}\sys{R}$.
        The conditional state of S is
        $\condstate{\superop\sys{RS}(\subsp{M}\sys{RS})}{\subsp{E} \sys{R}}$.
\end{itemize}
Intuitively, if R and S are completely independent (and perhaps widely
separated), the final S state should be independent of whether the 
measurement on R is performed before or after the evolution of S.
This is exactly the requirement for a Type E map.

\subsection{Representation theorem for MQT}

We now prove a result analogous to the representation theorem for
AQT.  Formally, we will show that
\begin{quotation}
    \noindent
    {\bf Representation theorem for generalized dynamics in MQT.}  
    Let S be a modal quantum system and $\superop$ be a Type M map 
    on subspaces for F.  Then the following conditions are equivalent.
    \begin{description}
        \item[(a)]  $\superop$ is Type E; that is, it can be extended
            in a way that commutes with the conditional operation.
        \item[(b)]  $\superop$ is Type I; that is, it can be expressed
            as invertible linear evolution on a larger system.
        \item[(c)]  $\superop$ is Type L; that is, it can be expressed
            as a mixture of linear maps satisfying Equation~\ref{MQToperatorsumnormalization}.
     \end{description}
\end{quotation}
As with the AQT result, this theorem is a strong characterization
of the ``reasonable'' evolution maps in modal quantum theory.  We
have argued that all reasonable maps are Type E, and any Type I map
is realizable by familiar linear evolution.  Constructive and
axiomatic approaches coincide.

We will prove the equivalence of the three conditions by establishing
the cyclic implication $\mbox{L} \Rightarrow \mbox{I} \Rightarrow \mbox{E}
\Rightarrow \mbox{L}$.  The first two implications are straightforward;
the last requires a bit more work.
Throughout, we take $\superop\sys{S}$ to be a Type M map on S.  

\bigskip

{$\mathbf{L \Rightarrow I}$:}
First, assume that $\superop\sys{S}$ is Type L.  This means that there
is a set of linear operators $\{ A_{k} \}$ that yield $\superop\sys{S}$
according to Equation~\ref{MQTKrausrep}, and that these operators satisfy
the normalization requirement (Equation~\ref{MQToperatorsumnormalization}).  
Now we introduce an environment system E whose dimension is equal to the 
number of $A_{k}$ operators.  We fix an initial E-state 
$\mket{0\sys{E}}$ and a basis $\{ \mket{k\sys{E}} \}$.

The set of SE states of the form $\mket{\phi\sys{S},0\sys{E}}$ constitute
a subspace.  Define the operator $T\sys{SE}$ on this subspace by
\begin{equation}
    T\sys{SE} \mket{\phi\sys{S},0\sys{E}} = \sum_{k} A_{k} \mket{\phi\sys{E}} \otimes \mket{k\sys{E}} .
\end{equation}
Because of the normalization requirement on the $A_k$ operators,
the right-hand side is never zero.  Thus, the operator $T\sys{SE}$
is one-to-one on the subspace, and so we may extend it to an
invertible operator on the whole of $\modalspace\sys{SE}$.  Given a 
mixed state $\subsp{M}\sys{S}$, it is straightforward to show
that
\begin{equation}
    \reduction{E} \left ( T\sys{SE}({\subsp{M}}\sys{S} \otimes \subsp{M}_{0}\sys{E} ) \right )
        = \vspan{ \{ A_{k} \subsp{M}\sys{S} \} }
        = \superop\sys{S} ( \subsp{M}\sys{S} ) ,
\end{equation}
where $\subsp{M}_{0}\sys{E} = \vspan{ \mket{0\sys{E}} }$.
The map $\superop\sys{S}$ is therefore Type I.

\bigskip

{$\mathbf{I \Rightarrow E}$:}
Now assume that $\superop\sys{S}$ is Type I, so that it is given by
invertible evolution on the extended system SE as above.   
For any additional system R we define the map
 \begin{equation}
   \superop \sys{RS} (\subsp{M}\sys{RS}) =  
   \reduction{E} \left [ (1\sys{R} \otimes T\sys{SE} )(\subsp{M}\sys{RS} \otimes {\subsp{M}}_0\sys{E}) \right ] .
\end{equation}  
As we have already remarked, the reduction operation
$\reduction{E}$ is an ``unconditional'' conditioning operation.
A direct application of the definition in Equation~\ref{condstatedef}
shows that iterated reduction with respect to independent subsystems
(in our case, R and E) can be done in any order.  This establishes that
every Type I map must also be Type E.

\bigskip
{$\mathbf{E \Rightarrow L}$:}
It only remains to prove that Type E implies Type L. Let $\superop\sys{S}$ 
be a Type E  map for a modal quantum system $S$, which is represented 
by a vector space of finite dimension $\dim \modalspace\sys{S} = d$.  
We append an identical quantum system R and consider the maximally
entangled state
\begin{equation}\label{maximallyentangledRSstate}
    \mket{\Phi\sys{RS}} =  \sum_{k} 
        \mket{k\sys{R},k\sys{S}} .
\end{equation}
Any initial state $\mket{\psi\sys{S}}$ of S could arise
in the following way.  The system RS is initially in the
entangled state $\mket{\Phi\sys{RS}}$, and then a measurement
is performed in R.  The resulting state of S, conditional on
the particular measurement outcome for R, happens to be 
$\mket{\psi\sys{S}}$.

We can do this more explicitly.  Given 
$\mket{\psi\sys{S}} = \sum_k g_{k} \mket{k\sys{S}}$,
we can construct the R-effect
\begin{equation}
     \mbra{\tilde{\psi}\sys{R}} = \sum_{k} {g_{k}}
       \mbra{k\sys{R}} .
\end{equation}
If $\mbra{\tilde{\psi}\sys{R}}$ corresponds to one outcome of a
basic measurement on R, then the 
associated conditional state of Q is $\mket{\psi\sys{Q}}$.

This reasoning can be generalized to mixed states. First, note that 
$\mathcal{M}\sys{RS}=\vspan{  \mket{\Phi\sys{RS}}  }$ 
is the one-dimensional mixed state that corresponds to the 
``fully entangled'' state $\mket{\Phi\sys{RS}}$. 
Now consider a general mixed Q-state 
\begin{equation}
\subsp{G}\sys{Q}= \vspan{\left \{ \mket{g\sys{Q}}=\sum_k g_{k} \mket{k\sys{S}} : (g_k)  \in G\right\}},
\end{equation}
where $G$ is a set of $d$-tuples $(g_k)$ of elements of $\scalarfield$.
Now define the R-effect
\begin{equation}
       \Gamma\sys{R}= \vspan{ \left\{ \mbra{g\sys{R}}=\sum_k g_{k} \mbra{k\sys{R}} : (g_k)  \in G\right\}}.
\end{equation}
Then
\begin{equation}
\subsp{G}\sys{Q} = \condstate{\mathcal{M}\sys{RS}}{\Gamma\sys{R}},
\end{equation}
since $\mket{g\sys{Q}} = \mamp{g\sys{R}}{\Phi\sys{RS}}$.

We use this machinery to construct a Type L representation the Type E map $\superop\sys{S}$. 
Let $\superop\sys{RS}$ be an extension of $\superop\sys{S}$. 
Let $\left\{ \mket{m_{\lambda}\sys{RS}} \right\}$ be a set of RS states 
such that $\superop\sys{RS}(\mathcal{M}\sys{RS}) = \vspan{\left\{ \mket{m_{\lambda}\sys{RS}} \right\}}$ 
(where $\lambda$ runs over some index set $\Lambda$).
Given states $\mket{g\sys{S}}$ and associated effects $\mbra{g\sys{R}}$ 
as described above, we define $A_\lambda$ as follows: 
\begin{equation}
A_{\lambda}\sys{S} \mket{g\sys{S}} = \mamp{g\sys{R}}{m_{\lambda}\sys{RS}}.
\end{equation}
We now have
\begin{eqnarray}
    \superop\sys{S}(\subsp{G}\sys{S}) 
                    &=& \superop\sys{S}(\condstate{\mathcal{M}\sys{RS}}{\Gamma\sys{R}}) \nonumber \\
                    &=& \condstate{\superop\sys{RS}(\mathcal{M}\sys{RS})}{\Gamma\sys{R}}\nonumber \\
                      &=& \condstate{\vspan{ \{ m_{\lambda}\sys{RS} \} }}{\Gamma\sys{R}} \nonumber \\
                 &=&  \vspan{\left\{ \mamp{g\sys{R}}{m_{\lambda}\sys{RS}} \right \} } \nonumber \\
                 &=&  \vspan{\left\{ {A_\lambda}\sys{S}\mket{g\sys{S}} \right\}  } \nonumber \\
     \superop\sys{S}(\subsp{G}\sys{S}) 
                 &=&  \bigvee_{\lambda}{A_\lambda}\sys{S}(\subsp{G}\sys{S}).
\end{eqnarray}
Thus, any Type E map is also Type L.

\bigskip

We see that Type E maps in MQT play a role parallel to CP maps in
actual quantum theory.  In fact, the connection is stronger than
this.  We can adapt the definition of Type E maps to AQT:  A linear
map $\superop\sys{Q}$ on density operators for Q is Type E provided
there exists an extended map $\superop\sys{RQ}$ on density operators
of RQ that commutes with the formation of conditional Q states from
effects on R.  It is not hard to show that this condition is 
{\em equivalent} to complete positivity of $\superop\sys{Q}$ 
and thus implies the existence of unitary (``Type L'') and Kraus (``Type L'') 
representations in AQT.


With the (now finished) proof of the representation theorem, we have
completed our development of modal quantum theory to include
generalized states, measurements, and dynamical evolution.  This
development has included both constructive approaches (based on the
basic axioms in Table~\ref{AQTandMQT}) and axiomatic characterizations.
The two routes lead to the same place, a fact that gives us confidence that
we have arrived at a ``complete'' development of the theory.
Further generalization will necessarily involve an extension 
of MQT to a more general type of theory.

\section{Generalized modal theories}

\subsection{Possibility tables for two systems}

In the study of the conceptual foundations of actual quantum 
theory, it is useful to consider AQT as an example of a more
general class of probabilistic theories.  Consider a system comprising
two subsystems, designated 1 and 2.  We can choose to make any of 
several possible measurements on each system and obtain various
joint results with various probabilities.  That is, our theory
allows us to compute probabilities of the form $p(x,y|X\sys{1},Y\sys{2})$,
the probability of the joint outcome $(x,y)$ given the choice of
measurement $X\sys{1}$ on system 1 and $Y\sys{2}$ on system 2.

The state of the composite system can thus be described by a
collection of joint probability distributions.  These may be
organized as a table.  The rows and columns of the table correspond
to the possible measurements on systems 1 and 2, respectively, like so:
\begin{equation} \label{generaltable}
\mbox{
\begin{tabular}{cccc}
    & $U\sys{2}$ & $V\sys{2}$ & $\cdots$ \\[1ex]
    $U\sys{1}$ &
   \fbox{\begin{tabular}{c|c|c}  &  & \\ \hline  & & \\ \hline & & \end{tabular}} &
   \fbox{\begin{tabular}{c|c|c}  &  & \\ \hline  & & \\ \hline & & \end{tabular}} & $\cdots$ \\[4ex]
   $V\sys{1}$ &
   \fbox{\begin{tabular}{c|c|c}  &  & \\ \hline  & & \\ \hline & & \end{tabular}} &
   \fbox{\begin{tabular}{c|c|c}  &  & \\ \hline  & & \\ \hline & & \end{tabular}} & $\cdots$ \\
   $\vdots$ & $\vdots$ & \vdots & 
   \end{tabular}}
\end{equation}
The theory is characterized by the set of possible states---that is,
the possible collections of distributions in the table.\footnote{AQT also
allows ``entangled'' measurements on composite systems, 
measurements which cannot be
reduced to separate measurements on the subsystems.  Probabilities for
non-entangled measurements, however, are sufficient to characterize 
the joint state of the system, so we restrict our attention to those.}

All of the tables we consider satisfy the {\em no-signalling principle}
which can be stated as follows \cite{gisin}.  For any choice of measurements
$A\sys{1}$, $B\sys{1}$, $C\sys{2}$ and $D\sys{2}$,
\begin{equation}
    \begin{array}{rcrcl}
    p\left(a|A\sys{1}\right) & = & 
    \sum_{c} p\left(a,c|A\sys{1},C\sys{2}\right) & = & 
    \sum_{d} p\left(a,d|A\sys{1},D\sys{2}\right) \\[1ex]
    p\left(c|C\sys{1}\right) & = & 
    \sum_{a} p\left(a,c|A\sys{1},C\sys{2}\right) & = &
    \sum_{b} p\left(b,c|B\sys{1},C\sys{2}\right) .
    \end{array}
\end{equation}
That is, the choice of system 2 measurement does not affect the
overall probability of a system 1 outcome, and {\em vice versa}.
In the table of joint distributions in Equation~\ref{generaltable},
this means that any two distributions in the same row are 
connected, in that their sub-rows sum to the same values.
A similar connection exists within each column as well.

Collections of distributions arising from a composite system in AQT
satisfy the no-signalling principle.  However, there are tables
satisfying this principle that could not arise in AQT.  These
include examples of ``states'' of a composite system that are
``more entangled'' than quantum mechanics allows \cite{prbox}.

We can adapt this approach to construct generalized modal theories 
that extend MQT.
The state of a composite system in a general modal theory would
be a table similar to the one in Equation~\ref{generaltable},
except that the the individual ``distributions'' only indicate
which joint outcomes are possible.  We use the
symbol $\myex$ to denote a possible outcome, and a blank space
for an impossible outcome.  For instance, the table 
for a pair of mobits in \zeetwo-MQT has just three rows and 
columns, corresponding to the three possible basic measurements
for each mobit.  For the entangled modal state 
$\mket{S} = \mket{0,1}-\mket{1,0}$, we have
\begin{equation} \label{mqtsinglettable}
\mathcal{S} = \left [
\mbox{
\begin{tabular}{cccc}
    & $X\sys{2}$ & $Y\sys{2}$ & $Z\sys{2}$ \\[1ex]
    $X\sys{1}$ &
      \fbox{\begin{tabular}{c|c}   & \myex \\ \hline \myex &   \end{tabular}} &
      \fbox{\begin{tabular}{c|c} \myex &   \\ \hline \myex & \myex \end{tabular}} &
      \fbox{\begin{tabular}{c|c} \myex & \myex \\ \hline   & \myex \end{tabular}} \\[4ex]
    $Y\sys{1}$ &
      \fbox{\begin{tabular}{c|c} \myex & \myex \\ \hline   & \myex \end{tabular}} &
      \fbox{\begin{tabular}{c|c}   & \myex \\ \hline \myex &   \end{tabular}} &
      \fbox{\begin{tabular}{c|c} \myex &   \\ \hline \myex & \myex \end{tabular}} \\[4ex]
    $Z\sys{1}$ &
      \fbox{\begin{tabular}{c|c} \myex &   \\ \hline \myex & \myex \end{tabular}} &
      \fbox{\begin{tabular}{c|c} \myex & \myex \\ \hline   & \myex \end{tabular}} &
      \fbox{\begin{tabular}{c|c}   & \myex \\ \hline \myex &   \end{tabular}}   
\end{tabular}} \right ]
\end{equation}

This table, like all tables arising from MQT systems, satisfies
a modal version of the no-signalling principle.  The question of
whether a particular subsystem result is possible does not depend
on what measurement is chosen for the other subsystem.  Thus, if
an $\myex$ occurs in the table, at least one $\myex$ must occur in
the corresponding sub-rows to the right and left, and in the 
corresponding sub-columns above and below.  We will only consider
general modal theories satisfying the modal no-signalling principle.

In a general probabilistic theory, we can take the convex combination
of two states and derive a ``mixed'' state.  The set of allowed
states is therefore a convex set.  In a general modal theory, 
the mixture of two tables $\mathcal{R}$
and $\mathcal{T}$ is simply $\mathcal{R} \vee \mathcal{T}$, the 
table in which a joint outcome is possible if it is possible in
either $\mathcal{R}$ or $\mathcal{T}$.  (This corresponds to the
usual mixture of states in MQT.)  Finally, there is a natural
partial ordering on states in a general modal theory.  We say that
$\mathcal{R} \preceq \mathcal{T}$ provided every possible result
in $\mathcal{R}$ is also possible in $\mathcal{T}$.

\subsection{Popescu-Rohrlich boxes}  \label{subsec:PRboxes}

Every generalized probabilistic table can be converted into a 
generalized modal table by replacing non-zero probabilities with
$\myex$  and zero probabilities with blanks.  A table obeying the
probabilistic no-signalling principle automatically yields one
that obeys the modal version.

We use this idea to create a modal version of an important 
example of a generalized probabilistic model, the 
``nonlocal box'' proposed by Popescu and Rohrlich \cite{prbox}.
This {\em PR box} satisfies the no-signalling principle
but is in a sense more entangled than allowed by AQT.
The modal version $\mathcal{P}$ looks like this:
\begin{equation} \label{modalPRbox}
\mathcal{P} = \left [
\mbox{
\begin{tabular}{ccc}
    & $C\sys{2}$ & $D\sys{2}$ \\[1ex]
    $A\sys{1}$ &
      \fbox{\begin{tabular}{c|c} \myex  &  \\ \hline  & \myex  \end{tabular}} &
      \fbox{\begin{tabular}{c|c} \myex  &  \\ \hline  & \myex  \end{tabular}} \\[4ex]
    $B\sys{1}$ &
      \fbox{\begin{tabular}{c|c} \myex  &  \\ \hline  & \myex  \end{tabular}} &
      \fbox{\begin{tabular}{c|c}   & \myex \\ \hline  \myex &  \end{tabular}}    
\end{tabular}} \right ]
\end{equation}
We can summarize this pattern of possibilities in a simple way:
For the measurement combinations $(A,C)$, 
$(A,D)$ and $(B,C)$ 
the joint measurement results must always agree, 
but for $(B,D)$ they always disagree.  
(The probabilistic PR box replaces $\myex$ with probability 1/2 in Equation~\ref{modalPRbox}.)

The PR box table $\mathcal{P}$ is minimal.  That is, if any
table $\mathcal{R}$ of similar dimensions satisfies the 
no-signalling principle, and if 
$\mathcal{R} \preceq \mathcal{P}$, 
then $\mathcal{R} = \mathcal{P}$.

Could the PR box table $\mathcal{P}$ in Equation~\ref{modalPRbox} 
arise from a composite system described by MQT?  In fact, it
cannot.  Since $\mathcal{P}$ is minimal, 
it suffices to consider only pure states
for system 12 together with measurements having
non-overlapping effects.  That is, the measurement $A\sys{1}$
consists of two effects (subspaces of $\dual{\modalspace\sys{1}}$)
$\subsp{A}_{+}$ and $\subsp{A}_{-}$ such that 
$\subsp{A}_{+} \cap \subsp{A}_{-} = \vspan{0}$, and so on.

Suppose $\mket{\Psi}$ is a modal quantum state that leads
to the PR box table $\mathcal{P}$ in Equation~\ref{modalPRbox}.
As shown in the Appendix, the upper-left quarter of the table
tells us that 
\begin{equation}
    \mket{\Psi} = \mket{\Psi_{+}} + \mket{\Psi_{-}},
\end{equation}
where these two non-zero parts of $\mket{\Psi}$ satisfy
\begin{eqnarray} \label{Psi12decomposition}
\mket{\Psi_{+}} \in \annihilator{\subsp{A}_{-}} \otimes \annihilator{\subsp{C}_{-}}
    & \mbox{and} &
    \mket{\Psi_{-}} \in \annihilator{\subsp{A}_{+}} \otimes \annihilator{\subsp{C}_{+}}.
\end{eqnarray}
The same state vector $\mket{\Psi}$ gives rise to the 
possibilities in the upper-right quarter of $\mathcal{P}$ also.
From this we can conclude that 
\begin{eqnarray}
\mket{\Psi_{+}} \in \annihilator{\subsp{A}_{-}} \otimes \annihilator{\subsp{D}_{-}}
    & \mbox{and} &
    \mket{\Psi_{-}} \in \annihilator{\subsp{A}_{+}} \otimes \annihilator{\subsp{D}_{+}}.
\end{eqnarray}
We can continue around the table $\mathcal{P}$, arriving at the 
following facts:
\begin{eqnarray}
\mket{\Psi_{+}} \in \annihilator{\subsp{B}_{+}} \otimes \annihilator{\subsp{D}_{-}}
    & \mbox{and} &
    \mket{\Psi_{-}} \in \annihilator{\subsp{B}_{-}} \otimes \annihilator{\subsp{D}_{+}}. \\
\mket{\Psi_{+}} \in \annihilator{\subsp{B}_{+}} \otimes \annihilator{\subsp{C}_{+}}
    & \mbox{and} &
    \mket{\Psi_{-}} \in \annihilator{\subsp{B}_{-}} \otimes \annihilator{\subsp{C}_{-}}.
\end{eqnarray}
This last pair of statements allows us to return to the
upper-left corner, concluding that
\begin{eqnarray}
\mket{\Psi_{+}} \in \annihilator{\subsp{A}_{+}} \otimes \annihilator{\subsp{C}_{+}}
    & \mbox{and} &
    \mket{\Psi_{-}} \in \annihilator{\subsp{A}_{-}} \otimes \annihilator{\subsp{C}_{-}}.
\end{eqnarray}
Since the annihilator subspaces are non-overlapping, this contradicts
Equation~\ref{Psi12decomposition}.  Thus, no such $\mket{\Psi}$ 
exists for which the set of possible measurement results is described 
by the PR box pattern $\mathcal{P}$.

\subsection{Probabilistic resolutions} \label{Probabilistic resolutions}

We have already noted that we can derive a generalized modal table
from a generalized probability table, while respecting the 
no-signalling principle.  Is it possible to do the reverse?
That is, if we have a table of possibilities for a modal system,
can we find a corresponding table of probabilities?  We call
this a {\em probabilistic resolution} of the modal table,
and distinguish two different types.
\begin{itemize}
    \item A {\em strong probabilistic resolution} assigns 
        zero probability to every impossible result and
        non-zero probability to every possible result ($\myex$).
    \item A {\em weak probabilistic resolution} assigns
        zero probability to every impossible result.  However,
        a ``possible'' result ($\myex$) may be assigned any
        probability, zero or non-zero.  (See the discussion
        in Subsection~\ref{subsec:modalworld}.)
\end{itemize}
In either case, we require that the resulting table of
distributions must satisfy the probabilistic no-signalling
principle.

As an example, consider the PR box table $\mathcal{P}$ of
Equation~\ref{modalPRbox}.  It is not difficult to show
that this table has only one allowed probabilistic resolution,
which is of the strong type:
\begin{equation} \label{modalPRboxresolution}
\mbox{
\begin{tabular}{ccc}
    & $A\sys{2}$ & $B\sys{2}$ \\[1ex]
    $A\sys{1}$ &
      \fbox{\begin{tabular}{c|c} 1/2  & 0 \\ \hline  0 & 1/2  \end{tabular}} &
      \fbox{\begin{tabular}{c|c} 1/2  & 0 \\ \hline  0 & 1/2  \end{tabular}} \\[4ex]
    $B\sys{1}$ &
      \fbox{\begin{tabular}{c|c} 1/2  & 0 \\ \hline  0 & 1/2  \end{tabular}} &
      \fbox{\begin{tabular}{c|c} 0 & 1/2 \\ \hline  1/2 & 0 \end{tabular}}    
\end{tabular}}
\end{equation}

Not all general modal tables actually have probabilistic resolutions
of either type.  Consider the following table (of which we have only
shown the relevant parts):
\begin{equation}  \label{NSPnotWPR}
\mathcal{N} = \left [
\mbox{
\begin{tabular}{cccc}
    & $U\sys{2}$ & $V\sys{2}$ & $W\sys{2}$ \\[1ex]
    $U\sys{1}$ &
      \fbox{\begin{tabular}{c|c|c}  & \myex &  \\ \hline  &  & \myex \\ \hline \myex &  &  \end{tabular}} &
      \fbox{\begin{tabular}{c|c|c} \myex &  &  \\ \hline  & \myex &  \\ \hline  &  & \myex \end{tabular}} &
       \\[4ex]
    $V\sys{1}$ &
      \fbox{\begin{tabular}{c|c|c} \myex &  &  \\ \hline  & \myex &  \\ \hline  &  & \myex \end{tabular}} &
      \fbox{\begin{tabular}{c|c|c} \myex &  &  \\ \hline  & \myex &  \\ \hline  &  & \myex \end{tabular}} &
      \fbox{\begin{tabular}{c|c|c} \myex &  &  \\ \hline  & \myex &  \\ \hline  & \myex &  \end{tabular}} \\[4ex]
    $W\sys{1}$ &
       &
      \fbox{\begin{tabular}{c|c|c} \myex & \myex &  \\ \hline  &  & \myex \\ \hline  &  &  \end{tabular}} &
      \fbox{\begin{tabular}{c|c|c} \myex &  &  \\ \hline  & \myex &  \\ \hline  &  &  \end{tabular}}   
\end{tabular}} \right ]
\end{equation}
By inspection, $\mathcal{N}$ satisfies the modal no-signalling 
principle.  When we attempt a probabilistic resolution, we quickly
discover that all of the possibilities in the $(U\sys{1},U\sys{2})$,
$(U\sys{1},V\sys{2})$, $(V\sys{1},U\sys{2})$ and $(V\sys{1},V\sys{2})$
sub-tables must be assigned probability 1/3.  We obtain
\begin{equation}
\mbox{
\begin{tabular}{cccc}
    & $U\sys{2}$ & $V\sys{2}$ & $W\sys{2}$ \\[1ex]
    $U\sys{1}$ &
      \fbox{\begin{tabular}{c|c|c}  & 1/3 &  \\ \hline  &  & 1/3 \\ \hline 1/3 &  &  \end{tabular}} &
      \fbox{\begin{tabular}{c|c|c} 1/3 &  &  \\ \hline  & 1/3 &  \\ \hline  &  & 1/3 \end{tabular}} &
       \\[4ex]
    $V\sys{1}$ &
      \fbox{\begin{tabular}{c|c|c} 1/3 &  &  \\ \hline  & 1/3 &  \\ \hline  &  & 1/3 \end{tabular}} &
      \fbox{\begin{tabular}{c|c|c} 1/3 &  &  \\ \hline  & 1/3 &  \\ \hline  &  & 1/3 \end{tabular}} &
      \fbox{\begin{tabular}{c|c|c} 1/3 &  & \phantom{1/3} \\ \hline  & 1/3 &  \\ \hline  & 1/3 &  \end{tabular}} \\[4ex]
    $W\sys{1}$ &
       &
      \fbox{\begin{tabular}{c|c|c} 1/3 & 1/3 &  \\ \hline  &  & 1/3 \\ \hline  & \phantom{1/3} &  \end{tabular}} &
      \fbox{\begin{tabular}{c|c|c} $p$ &  &  \\ \hline  & $q$ &  \\ \hline \phantom{1/3} & \phantom{1/3} & \phantom{1/3} \end{tabular}}   
\end{tabular}}
\end{equation}
where we have for clarity omitted zero entries.
The trouble arises in the lower-right corner $(W\sys{1},W\sys{2})$.  
The probabilistic no-signalling principle imposes two sets 
of constraints on the probabilities $p$ and $q$.  
Comparing to the $(W\sys{1},V\sys{2})$ sub-table,
we require $p = 2/3$ and $q = 1/3$.  Comparing to the
$(V\sys{1},W\sys{2})$ sub-table, we require $p = 1/3$ and $q = 2/3$.
We therefore conclude that no probabilistic resolution exists
for modal table $\mathcal{N}$.

Under some circumstances, we can guarantee that a probabilistic
resolution must exist.  Suppose that a general modal table $\mathcal{R}$
arises from a local hidden variable theory.  For each particular
value $h$ of the hidden variables, the outcomes of all joint
measurements are determined.  The resulting table $\mathcal{D}_{h}$
is thus deterministic---that is, each sub-table contains only a single
$\myex$.  The overall table $\mathcal{R}$ is thus a mixture of
different $\mathcal{D}_{h}$ tables.  Locality
of the hidden variable theory means that each deterministic table
$\mathcal{D}_{h}$ individually satisfies the no-signalling
principle.

Suppose there are $N$ distinct deterministic tables
$\mathcal{D}_{h}$.  Each deterministic table has an obvious
probabilistic resolution in which each $\myex$ entry is
given probability 1.  Now we assign each distinct
$\mathcal{D}_{h}$ a probability of $1/N$, and
take a mixture of their probabilistic resolutions
with these weights.  That is, if a particular outcome
of a particular joint measurement is possible in 
$M$ of the deterministic tables, it is assigned 
an overall probability $M/N$.  The resulting table
satisfies the probabilistic no-signalling principle,
since it is a convex combination of no-signalling tables.
Furthermore, it is a strong probabilistic resolution
of $\mathcal{R}$, since it assigns a probability
at least $1/N$ to each possible measurement outcome.
Therefore, every general modal table arising from a
local hidden variable theory has a strong probabilistic
resolution.

The converse is certainly false.  The PR box table
$\mathcal{P}$ of Equation~\ref{modalPRbox} has a
strong probabilistic resolution (Equation~\ref{modalPRboxresolution}).
However, $\mathcal{P}$ is a minimal table, which
means it cannot arise as a mixture of deterministic
tables that satisfy the no-signalling principle.
Therefore $\mathcal{P}$ cannot arise from any
local hidden variable theory.

Now consider the modal table $\mathcal{S}$ arising from
the \zeetwo-MQT singlet state, as shown in
Equation~\ref{mqtsinglettable}.  This has a unique
probabilistic resolution, which we display below.
Note that some of the possible outcomes have to
be assigned probability zero---that is, only a
weak probabilistic resolution can be given for this table:
\begin{equation} \label{mqtsingletresolution}
\mbox{
\begin{tabular}{cccc}
    & $X\sys{2}$ & $Y\sys{2}$ & $Z\sys{2}$ \\[1ex]
    $X\sys{1}$ &
      \fbox{\begin{tabular}{c|c}   & 1/2 \\ \hline 1/2 &   \end{tabular}} &
      \fbox{\begin{tabular}{c|c} 1/2 &   \\ \hline 0 & 1/2 \end{tabular}} &
      \fbox{\begin{tabular}{c|c} 1/2 & 0 \\ \hline   & 1/2 \end{tabular}} \\[4ex]
    $Y\sys{1}$ &
      \fbox{\begin{tabular}{c|c} 1/2 & 0 \\ \hline   & 1/2 \end{tabular}} &
      \fbox{\begin{tabular}{c|c}   & 1/2 \\ \hline 1/2 &   \end{tabular}} &
      \fbox{\begin{tabular}{c|c} 1/2 &   \\ \hline 0 & 1/2 \end{tabular}} \\[4ex]
    $Z\sys{1}$ &
      \fbox{\begin{tabular}{c|c} 1/2 &   \\ \hline 0 & 1/2 \end{tabular}} &
      \fbox{\begin{tabular}{c|c} 1/2 & 0 \\ \hline   & 1/2 \end{tabular}} &
      \fbox{\begin{tabular}{c|c}   & 1/2 \\ \hline 1/2 &   \end{tabular}}   
\end{tabular}}
\end{equation}

There are a number of things to remark about the probabilistic
resolution in Equation~\ref{mqtsingletresolution}.  The MQT
singlet state $\mket{S}$ gives us an example of a table with
a weak probabilistic resolution but not a strong probabilistic
resolution.  This gives us another proof that the modal properties
of $\mket{S}$ (represented in table $\mathcal{S}$) cannot be derived 
from any local hidden variable theory: if such a theory existed,
the table would certainly have a strong probabilistic resolution.

As we have seen, the modal PR box table $\mathcal{P}$ 
of Equation~\ref{modalPRbox} cannot arise from an entangled 
composite system in MQT.  Nevertheless, the
weak probabilistic resolution of Equation~\ref{mqtsingletresolution}
does contain a {\em probabilistic} PR box!  Consider the 
following section of the table:
\begin{equation}
\mbox{
\begin{tabular}{ccc}
    & $Z\sys{2}$ & $Y\sys{2}$ \\[1ex]
    $X\sys{1}$ &
      \fbox{\begin{tabular}{c|c} 1/2  & 0 \\ \hline  0 & 1/2  \end{tabular}} &
      \fbox{\begin{tabular}{c|c} 1/2  & 0 \\ \hline  0 & 1/2  \end{tabular}} \\[4ex]
    $Y\sys{1}$ &
      \fbox{\begin{tabular}{c|c} 1/2  & 0 \\ \hline  0 & 1/2  \end{tabular}} &
      \fbox{\begin{tabular}{c|c} 0 & 1/2 \\ \hline  1/2 & 0 \end{tabular}}    
\end{tabular}}
\end{equation}
This apparent paradox arises because a weak probabilistic resolution
allows probability zero to be assigned to a possible measurement
outcome.

A probabilistic PR box cannot arise in actual quantum theory.  
It follows that the behavior of an entangled composite system 
in MQT cannot be ``simulated'' by an entangled composite system 
in AQT.  (This is why the pseudo-telepathy game for $\mket{S}$ 
described in Subsection~\ref{subsec:entanglement} has no winning 
strategy if the players can only share entangled states from AQT.)

\subsection{A hierarchy of modal theories}

We have considered several distinct types of two-system modal tables.
\begin{itemize}
    \item  NSP is the set of tables satisfying the no-signalling principle.
        (This is our ``universe'' of tables.)
    \item  SPR is the set of tables that have a strong probabilistic resolution.
    \item  WPR is the set of tables that have a weak probabilistic resolution.
    \item  LHV is the set of tables that have a local hidden variable model.
    \item  MQT is the set of tables that can arise from a bipartite system
        in modal quantum theory.
\end{itemize}
As we have seen there are several relations between these classes:
\begin{equation}
    \mbox{LHV} \subset \mbox{SPR} \subset \mbox{WPR} \subset \mbox{NSP}.
\end{equation}
The inclusion relation is strict in each case.  The PR box table 
$\mathcal{P}$ in in Equation~\ref{modalPRbox} is in SPR but not LHV;
the $\zeetwo$ modal singlet table $\mathcal{S}$ in Equation~\ref{mqtsinglettable} 
is in WPR but not SPR; and the table $\mathcal{N}$ in Equation~\ref{NSPnotWPR}
is in NSP but not WPR.

What about the set MQT?  It is not hard to see that every table
in LHV is also in MQT.  We also know there are tables that are
in MQT but not in LHV or SPR.  Conversely, the PR box $\mathcal{P}$
(Equation~\ref{modalPRbox}) is in SPR and WPR but not MQT.
It remains to pin down the relation between MQT and WPR.  
We will prove that $\mbox{MQT} \subset \mbox{WPR}$---that
is, that every table that arises from the state of 
a bipartite system in MQT must have a weak probabilistic resolution.

To establish this, we will take advantage of several simplifications.
Since a weak probabilistic resolution allows us to assign $p=0$
for some possible outcomes, the addition of possibilities ($\myex$ entries)
to a modal table can never frustrate a weak probabilistic resolution.
Therefore, we need only consider minimal modal tables in MQT, those
that arise from pure bipartite states.

Every pure bipartite state $\mket{\Psi}$ has a Schmidt decomposition 
(as in Equation~\ref{MQTSchmidtdecomposition}) with an integer
Schmidt number $s$.  The state vector therefore lies in 
a subspace we may denote $\modalspace \otimes \modalspace$, with
$\dim \modalspace = s$.  The space $\modalspace$ is a subspace
of the state spaces for the two systems; but we can regard it
as the effective state space for the particular situation described
by $\mket{\Psi}$.  Any measurement on either subsystem can hence
be regarded as a generalized measurement on $\modalspace$.  
Therefore, we can 
suppose that $\mket{\Psi}$ is a state of maximum Schmidt number 
for a pair of identical systems with state spaces $\modalspace$ of
dimension $s$.
(The case where $s = 1$ is trivial, so we will assume that
$s \geq 2$ and $\mket{\Psi}$ is entangled.)

Generalized measurements whose effect subspaces have 
$\dim \subsp{E}_{a} > 1$ can be viewed as ``coarse-grained'' 
versions of measurements with one-dimensional
(``fine-grained'') effects.  If we can construct a weak probabilistic 
resolution for the fine-grained measurements, this will automatically
give a resolution for the coarse-grained version.  Therefore, we need
only consider fine-grained measurements---that is, those whose effect
subspaces are one-dimensional.

A fine-grained measurement can be viewed as a spanning set for 
$\dual{\modalspace}$.  Every such spanning set contains a basis,
and at least one of these basis effects must be possible for
a given state.  The ``extra'' effects can always be assigned
probability zero.  Therefore, we need only consider basic measurements,
those that correspond to basis sets for $\dual{\modalspace}$.

Armed with all of these simplifications, let us consider a
pair of identical systems in a pure entangled state
$\mket{\Psi\sys{12}}$ of maximum Schmidt number.  For each pair
of basic measurements, we arrive at an $s \times s$ sub-table
of possibilities.  Let us focus our attention on one such
sub-table, with measurement bases $\{ \mbra{e_{j}\sys{1}} \}$
(the rows) and $\{ \mbra{f_{k}\sys{2}} \}$ (the columns).

For each $\mbra{e_{j}\sys{1}}$, define the set 
\begin{equation}
    F_{j} = \{ \mbra{f_{k}\sys{2}} : \mamp{e_{j}\sys{1} f_{k}\sys{2}}{\Psi\sys{12}} \neq 0 \} .
\end{equation}
That is, for each system 1 effect, we consider the set of
system 2 effects that are jointly possible 
given state $\mket{\Psi\sys{12}}$.  Consider next
a set $E$ containing $d$ system 1 effects $\mbra{e_{j}\sys{1}}$.
Each $\mbra{e_{j}\sys{1}}$ corresponds to a conditional
state $\mket{\psi_{j}\sys{2}} = \mamp{e_{j}\sys{1}}{\Psi\sys{12}}$.
Since $\mket{\Psi\sys{12}}$ is maximally entangled, these
are non-zero and linearly independent.  Hence, the 
effects in $E$ correspond to a set of system 2 states
that span a subspace $\subsp{M}_{E}\sys{2}$ of dimension $d$.

A basic system 2 measurement on $\subsp{M}_{E}$ must have
at least $d$ possible outcomes.  These correspond to the
system 2 effects in the set $\displaystyle \bigcup_{E} F_{j}$.
We have shown that the collection $F = \{ F_{j} \}$ of
sets has the property that, for any set $E$ of basic
system 1 effects,
\begin{equation}
    \# \left ( \bigcup_{E} F_{j} \right ) \geq \# \left ( E \right ), 
\end{equation}
where $\# ( K )$ is the number of elements in finite set $K$.
By Hall's Marriage Theorem \cite{hallmarriage}, we can conclude
that the collection $F$ has a set of distinct representatives.
That is, for each $\mbra{e_{j}\sys{1}}$ we can identify a
corresponding $\mbra{f_{j}\sys{2}}$ such that 
\begin{itemize}
    \item $\mamp{e_{j}\sys{1} f_{j}\sys{2}}{\Psi\sys{12}} \neq 0$ for all $j$, and
    \item $\mbra{f_{i}\sys{2}} \neq \mbra{f_{j}\sys{2}}$ when $i \neq j$.
\end{itemize}
In our sub-table, this means we can identify a set of the
possible joint outcomes (the $\myex$'s) such that each
row and each column contains exactly one of them.

We therefore make the following probability assignment.
Each impossible joint outcome, of course, is assigned $p=0$.
We also assign $p=0$ to all of the possible joint outcomes
except for those we have identified above, one in each
row and column.  These are assigned $p = 1/s$.

The same procedure can be applied for each sub-table 
independently.  In every case, the total probability
for each row and for each column is $1/s$.  Therefore,
the probabilistic no-signalling principle is automatically
satisfied.  Our construction (via Hall's Marriage Theorem)
yields a weak probabilistic resolution for the modal table
associated with the entangled state $\mket{\Psi\sys{12}}$.
Every table that arises from a bipartite state in MQT has
a weak probabilistic resolution.

\begin{figure} \label{fig:hierarchy}
\begin{center}
    \includegraphics[width=3.5in]{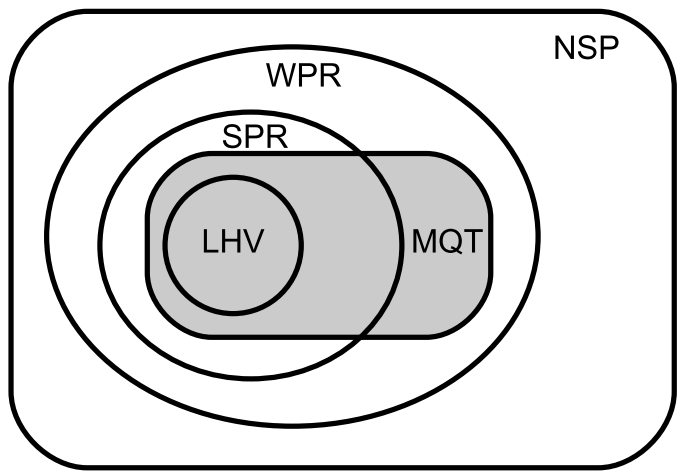}
\end{center}
\caption{The hierarchy of bipartite states in modal
    theories.}
\end{figure}

In terms of our hierarchy of modal theories, we have
shown that MQT $\subset$ WPR.  Our conclusions are
summarized in Figure~\ref{fig:hierarchy}.
It is worth noting that all of the six distinct regions
in this diagram are non-empty.  Thus, for example, 
table $\mathcal{S}$ of Equation~\ref{mqtsinglettable}
is in MQT but not SPR; table $\mathcal{P}$ of 
Equation~\ref{modalPRbox} is in SPR but not MQT; and
table $\mathcal{N}$ of Equation~\ref{NSPnotWPR} is
within NSP but not WPR.  Other examples are easy to
construct.

\section{Concluding Remarks}

\subsection{What MQT has, and what it does not have}

As diverting an exercise as MQT is, its real purpose is to
shed light on the structure of actual quantum theory.
It is remarkable how many of the features of AQT are 
retained, at least in some form, even in such a primitive theory.
An incomplete summary can be found in Figure~\ref{fig:MQTfeatures}.
\begin{figure}  \label{fig:MQTfeatures}
\begin{center}
\begin{tabular}{cc}
    \begin{minipage}[t]{2.5in}
    \noindent {\bf MQT does not have:}
    \begin{itemize}
    \item  Probabilities, expectations
    \item  ($\scalarfield$ finite) Continuous sets of states and
        observables, or continuous time evolution
    \item  Inner product, outer product, orthogonality
    \item  Convexity
    \item  Hermitian conjugation ($\dagger$)
    \item  Density operators
    \item  Effect operators
    \item  CP maps
    \item  Unextendable product bases
    \end{itemize}
    \end{minipage}
    &
    \begin{minipage}[t]{2.5in}
    \noindent {\bf MQT does have:}
    \begin{itemize} 
    \item  ``Classical'' versus ``quantum'' theories
    \item  Superposition, interference
    \item  Complementary measurements
    \item  Entanglement
    \item  No local hidden variables
    \item  Kochen-Specker theorem, ``free will'' theorem
    \item  Superdense coding, teleportation, ``steering'' of mixtures
    \item  Mixed states, generalized effects, generalized evolution maps
    \item  No cloning theorem
    \item  Nonclassical models of computation
    \end{itemize}
    \end{minipage}
\end{tabular}
\end{center}
\caption{Properties and structures of actual quantum theory that 
    either are or are not present in MQT.}
\end{figure}
In the left-hand column we have listed aspects of 
AQT that are not found in MQT; in the right-hand column,
we have listed aspects of AQT that do have analogies 
in MQT.  The key point is that {\em nothing
in the right-hand column logically depends on anything 
in the left-hand column.}

Furthermore, as we have seen, the process of generalization
is very similar in AQT and MQT.  In both theories we can
develop more general concepts of state, measurement and 
time evolution, and these generalizations can be characterized
in both constructive and axiomatic ways.  Both theories can
also be extended to more general (probabilistic or modal) theories.
Within these more general types of theories, the quantum theories
have special properties---e.g., PR boxes are excluded in either
theory, and every bipartite state in MQT has a weak probabilistic 
resolution.

This last point deserves further comment.  We have imagined a 
modal world, one which supports the distinction between 
``possible'' and ``impossible'' events without necessarily 
imposing any probability measure.  As we have seen, it is
not always possible to make a reasonable probability assignment
in such a modal world.  The table $\mathcal{N}$ of 
Equation~\ref{NSPnotWPR} provides an example that 
respects the modal no-signalling principle, 
but within which we cannot assign probabilities respecting
the probabilistic NSP.

Under what circumstances, then, can we make reasonable 
probability assignments to a set of possibilities?  In
the bipartite case, we have shown that this can always
be done for joint measurements on a modal quantum system.
That is, the underlying structure of MQT somehow 
``makes room'' for probabilities.  It remains to be seen
whether this sheds any light on the way in which 
probabilities arise in the real world.

\subsection{Open problems}

Modal quantum theory is an exceptionally rich ``toy model''
of physics.  Despite the known features of the theory
summarized in Figure~\ref{fig:MQTfeatures}, there remain
many open questions.
\begin{itemize}
\item  Although we have shown that bipartite systems 
    in MQT support weak probabilistic resolutions, 
    we do not know whether this is true for
    entangled states of three or more systems.
\item  We have established many properties of pure
    entangled states for MQT system, but we know
    much less about mixed entangled states.  For
    example, we do not know whether there are 
    ``bound'' entangled states in MQT \cite{boundentanglement}.  
    (The usual
    AQT construction cannot be adapted to MQT, since
    there are no unextendable product bases in MQT.)
\item  Many results and ideas of quantum information and
    quantum computation have direct analogues in MQT.
    For instance, MQT supports both superdense coding
    and teleportation \cite{mqtpaper}.  
    It is straightforward to show
    that the Deutsch-Jozsa oracle algorithm (distinguishing
    constant and balanced functions with a single query)
    can be implemented without change on a modal quantum
    computer with $\scalarfield = \mathbb{Z}_{3}$
    \cite{djalgorithm}.
    However, a great deal of work remains to be done
    along these lines.\footnote{Some observations are
    obvious.  In a world without probabilities, we are
    interested in the {\em zero-error capacities} of
    communication channels and computer algorithms that
    reach {\em deterministic} results.}
\item  It is possible to regard actual quantum theory
    as a special type of modal quantum theory in which
    $\scalarfield = \mathbb{C}$ and we have special
    restrictions on the allowed measurements and 
    time evolution operators.  What (if anything) can be
    gained by analyzing AQT in this way?
\end{itemize}
We believe that the investigation of these and other
open problems MQT will shed further light on the 
mathematical structure of quantum theory.

\section{Acknowledgments}

We have benefitted from discussions of MQT with many 
colleagues.  Howard Barnum and Alex Wilce helped us
clarify the mathematical representation of measurement
within the theory.
Charles Bennett and John Smolin suggested
several questions about entangled states.
Gilles Brassard pointed out that the ``no hidden variables''
results in MQT are best described by pseudo-telepathy games.
Our research students Arjun Singh (Denison) and Peter Johnson
(Kenyon) participated in the early development 
of the MQT model.
Rob Spekkens (no stranger to thought-provoking ``foil''
theories) has been particularly helpful at many stages
of this project.

We would also like to thank the Perimeter Institute for its
hospitality and the organizers of the workshop there on
``Conceptual Foundations and Foils for Quantum Information
Processing'', May 9--13, 2011.

\section*{Appendix}

Here we fill in the details of the argument in Subsection~\ref{subsec:PRboxes}.
For convenience, we will suppose that modal quantum systems 1 and 2
are both described by the state space $\modalspace$, and that the
same two-outcome measurement is performed on each.  The effect
subspaces $\subsp{E}$ and $\subsp{F}$ in $\dual{\modalspace}$
are non-overlapping, so that $\subsp{E} \cap \subsp{F} = \vspan{0}$.
Finally, we assume that the joint possibility table for the state
$\mket{\Psi}$ is as follows:
\begin{equation} \label{correlatedsubtable}
\begin{tabular}{cc}
        & \begin{tabular}{cc} $\subsp{E}$ & $\subsp{F}$ \end{tabular} \\[1ex]
    \begin{tabular}{c} $\subsp{E}$ \\[0.5ex] $\subsp{F}$ \end{tabular} & 
    \fbox{\begin{tabular}{c|c} \myex  &  \\ \hline  & \myex  \end{tabular}}
\end{tabular}
\end{equation}
(Each sub-table of Equation~\ref{modalPRbox} is of this form.)

We can find a basis for $\dual{\modalspace}$ of the form
$\{ \mbra{e_{i}}, \mbra{f_{m}} \}$, where the $\{ \mbra{e_{i}} \}$
spans $\subsp{E}$ and $\{ \mbra{f_{m}} \}$ spans $\subsp{F}$.
The dual basis $\{ \mket{e_{i}}, \mket{f_{m}} \}$ of $\modalspace$
therefore has the property that $\mket{e_{i}}$ is annihilated
by every $\mbra{f_{m}}$ and $\mket{f_{m}}$ is annihilated by
every $\mbra{e_{i}}$.  In fact, $\{ \mket{e_{i}} \}$ spans
the annihilator $\annihilator{\subsp{F}}$ and $\{ \mket{f_{m}} \}$
spans $\annihilator{\subsp{E}}$.  We can expand the composite
state $\mket{\Psi}$ in this way:
\begin{equation}
    \mket{\Psi} = \sum_{ij} \alpha_{ij} \mket{e_{i} e_{j}}
        + \sum_{in} \beta_{in} \mket{e_{i} f_{n}}
        + \sum_{mj} \gamma_{mj} \mket{f_{m} e_{j}}
        + \sum_{mn} \delta_{mn} \mket{f_{m} f_{n}} .
\end{equation}
From Equation~\ref{correlatedsubtable}, we can see that the
effect $\subsp{E} \otimes \subsp{F}$ is impossible, which implies
that $\mamp{e_{i} f_{n}}{\Psi} = \beta_{in} = 0$ for every $i,n$.
In the same way, because $\subsp{F} \otimes \subsp{E}$ is impossible,
$\gamma_{mj} = 0$ for every $m,j$.  Therefore,
\begin{equation}
    \mket{\Psi} = \mket{\Psi_{ee}} + \mket{\Psi_{ff}} ,
\end{equation}
where $\mket{\Psi_{ee}} \in \annihilator{\subsp{F}} \otimes \annihilator{\subsp{F}}$
and $\mket{\Psi_{ff}} \in \annihilator{\subsp{E}} \otimes \annihilator{\subsp{E}}$.

Though we have supposed that the two systems are of the same type and that
the same measurement is made on each, it is easy to adapt this argument
to more general situations, provided the effect subspaces are
non-overlapping.


\end{document}